%% file: ctv.tex
\documentclass[11pt]{article}
\pdfoutput=1 
\usepackage{mathtools}
\usepackage{booktabs}

\usepackage{amsmath,amssymb,amsbsy,amstext, amsthm, simplewick, amsfonts,braket}
\usepackage{graphicx}
\usepackage[small]{caption}
\usepackage{siunitx}
\usepackage{subcaption}
\usepackage{upgreek}
\usepackage{framed}
\usepackage{wrapfig}
\usepackage{multirow}
\usepackage{bbm}
\usepackage[numbers,sort&compress]{natbib}
\usepackage[svgnames,dvipsnames,x11names]{xcolor}
\usepackage[utf8x]{inputenc}
\usepackage{selinput} 
\usepackage{bm}
\usepackage{float}
\usepackage{dsfont}
\usepackage[margin=1cm]{caption}
\usepackage{subcaption}
\usepackage{sidecap}
\usepackage{longtable}
\usepackage{anyfontsize}

\usepackage{epstopdf}
\usepackage{cancel}
\usepackage{tcolorbox}
\usepackage{latexsym,amsmath,amssymb,epsfig}
\usepackage{braket}
\usepackage{tensor}
\usepackage{tocloft}
\usepackage[permil]{overpic}

\usepackage{tcolorbox}
\definecolor{greyish2}{rgb}{.96,.96,.96}

\def\xyma{\xymatrix@M.7em}
\def\xymas{\xymatrix@M.1em}

\newcommand{\Comment}[1]{{}}
\definecolor{darkblue}{rgb}{0.15,0.35,0.55}
\definecolor{reddish}{rgb}{0.65, 0.2, 0.2}
\definecolor{darkgreen}{RGB}{50,150,0}
\definecolor{greyish2}{rgb}{.96,.96,.96}
\usepackage[linktocpage=true]{hyperref}
\hypersetup{
colorlinks=true,
citecolor=darkblue,
linkcolor=darkblue,
urlcolor=darkblue,
pdfauthor={},
pdftitle={},
pdfsubject={}
}

\DeclareFontFamily{OT1}{rsfs10}{}
\DeclareFontShape{OT1}{rsfs10}{m}{n}{ <-> rsfs10 }{}
\DeclareMathAlphabet{\mathscript}{OT1}{rsfs10}{m}{n}
\DeclareMathAlphabet{\mathbbold}{U}{bbold}{m}{n}

\def\gsim{ \lower .75ex \hbox{$\sim$} \llap{\raise .27ex \hbox{$>$}} }
\def\lsim{ \lower .75ex \hbox{$\sim$} \llap{\raise .27ex \hbox{$<$}} }
\def\be{\begin{equation}}
\def\ee{\end{equation}}
\def\bea{\begin{eqnarray}}
\def\eea{\end{eqnarray}}

\newcommand{\baaa}{\begin{eqnarray}}
\newcommand{\eaaa}{\end{eqnarray}}

\newcommand{\para}[1]{\paragraph{#1}\mbox{}\\}

\usepackage{tikz}
\usetikzlibrary{decorations}
\usetikzlibrary{calc}
\pgfdeclaredecoration{complete sines}{initial}
{
    \state{initial}[
        width=+0pt,
        next state=upsine,
        persistent precomputation={\pgfmathsetmacro\matchinglength{
            \pgfdecoratedinputsegmentlength / int(\pgfdecoratedinputsegmentlength/\pgfdecorationsegmentlength)}
            \setlength{\pgfdecorationsegmentlength}{\matchinglength pt}
        }] {}
    \state{upsine}[width=\pgfdecorationsegmentlength,next state=downsine]{
        \pgfpathsine{\pgfpoint{0.25\pgfdecorationsegmentlength}{0.5\pgfdecorationsegmentamplitude}}
        \pgfpathcosine{\pgfpoint{0.25\pgfdecorationsegmentlength}{-0.5\pgfdecorationsegmentamplitude}}
    }
    \state{downsine}[width=\pgfdecorationsegmentlength,next state=upsine]{
        \pgfpathsine{\pgfpoint{0.25\pgfdecorationsegmentlength}{-0.5\pgfdecorationsegmentamplitude}}
        \pgfpathcosine{\pgfpoint{0.25\pgfdecorationsegmentlength}{0.5\pgfdecorationsegmentamplitude}}
}
    \state{final}{}
}

\definecolor{greyish}{rgb}{.90,.90,.90}
\definecolor{greyish2}{rgb}{.96,.96,.96}
\usepackage{xcolor,colortbl}
\usepackage{tcolorbox}

\usepackage[all]{xy}

\newcommand{\p}{\partial}

\DeclareSymbolFont{matha}{OML}{txmi}{m}{it}

\newcommand{\bh}{\bar{h}}

\newcommand{\cp}[1]{\vcenter{\hbox{#1}}}
\newcommand{\id}{\mathbbold{1}}
\newcommand{\braidB}{\mathbbold{B}}
\newcommand{\modularS}{\mathbb{S}}
\newcommand{\hatS}{\widehat{\mathbb{S}}}

\newcommand{\fusionF}[2]{
\mathbb{F}_{#1}
\begin{bsmallmatrix} #2 \end{bsmallmatrix}
}
\newcommand{\vbracket}[1]{\left\langle\cp{#1}\right\rangle}

\def\centerarc[#1](#2)(#3:#4:#5)
{ \draw[#1] ($(#2)+({#5*cos(#3)},{#5*sin(#3)})$) arc (#3:#4:#5); }

\usepackage{adjustbox}
\newcommand{\graphS}[5]{\cp{
\adjustbox{margin=-1.5ex 0 -1ex 0}{
\begin{tikzpicture}[scale=0.25]
\coordinate (upperLeft) at (0,1);
\coordinate (lowerLeft) at (0,-1);
\coordinate (vertexLeft) at (0.8,0);
\coordinate (upperRight) at (3.5,1);
\coordinate (lowerRight) at (3.5,-1);
\coordinate (vertexRight) at (2.7,0);
\draw (upperLeft) -- (vertexLeft);
\draw (lowerLeft) -- (vertexLeft);
\draw (vertexLeft) -- (vertexRight);
\draw (upperRight) -- (vertexRight);
\draw (lowerRight) -- (vertexRight);
\node[xshift=-3] at (upperLeft) {\footnotesize \ensuremath{#1}};
\node[xshift=-3] at (lowerLeft) {\footnotesize \ensuremath{#2}};
\node[xshift=3] at (upperRight) {\footnotesize \ensuremath{#3}};
\node[xshift=3] at (lowerRight) {\footnotesize \ensuremath{#4}};
\node at (1.8,0.7) {\footnotesize \ensuremath{#5}};
\end{tikzpicture}
}}}
\newcommand{\graphSbig}[5]{
\adjustbox{margin=-1.5ex 0 -1ex 0}{
\begin{tikzpicture}[scale=0.5]
\coordinate (upperLeft) at (0,1);
\coordinate (lowerLeft) at (0,-1);
\coordinate (vertexLeft) at (0.8,0);
\coordinate (upperRight) at (3.5,1);
\coordinate (lowerRight) at (3.5,-1);
\coordinate (vertexRight) at (2.7,0);
\draw[ultra thick] (upperLeft) -- (vertexLeft);
\draw[ultra thick] (lowerLeft) -- (vertexLeft);
\draw[ultra thick] (vertexLeft) -- (vertexRight);
\draw[ultra thick] (upperRight) -- (vertexRight);
\draw[ultra thick] (lowerRight) -- (vertexRight);
\node[xshift=-5] at (upperLeft) {\footnotesize \ensuremath{#1}};
\node[xshift=-5] at (lowerLeft) {\footnotesize \ensuremath{#2}};
\node[xshift=5] at (upperRight) {\footnotesize \ensuremath{#3}};
\node[xshift=5] at (lowerRight) {\footnotesize \ensuremath{#4}};
\node at (1.8,0.5) {\footnotesize \ensuremath{#5}};
\end{tikzpicture}
}}
\newcommand{\graphT}[5]{\cp{
\adjustbox{margin=-1.5ex 0 -1ex 0}{
\begin{tikzpicture}[scale=0.25]
\coordinate (upperLeft) at (-1,3.5);
\coordinate (lowerLeft) at (-1,0);
\coordinate (upperRight) at (1,3.5);
\coordinate (lowerRight) at (1,0);
\coordinate (vertexTop) at (0,2.7);
\coordinate (vertexBottom) at (0, .8);
\draw (upperLeft) -- (vertexTop);
\draw (upperRight) -- (vertexTop);
\draw (lowerLeft) -- (vertexBottom);
\draw (lowerRight) -- (vertexBottom);
\draw (vertexTop) -- (vertexBottom);
\node[xshift=-3] at (upperLeft) {\footnotesize \ensuremath #1};
\node[xshift=-3] at (lowerLeft) {\footnotesize \ensuremath #2};
\node[xshift=3] at (upperRight) {\footnotesize \ensuremath #3};
\node[xshift=3] at (lowerRight) {\footnotesize \ensuremath #4};
\node at (0.7,1.8) {\footnotesize \ensuremath #5};
\end{tikzpicture}
}}}

\newcommand{\torusBlockM}[3]{
\cp{
\adjustbox{margin=-1ex -1ex -1ex -1ex}{
\begin{tikzpicture}[scale=0.25]
\coordinate (circleCenter) at (0,2);
\coordinate (vertex) at (0,1);
\coordinate (bottom) at (0,0);
\draw (bottom)-- (vertex);
\draw (circleCenter) circle (1);
\node at (circleCenter) {\footnotesize #3};
\node[xshift=-6,yshift=2] at (bottom) {\footnotesize \ensuremath{#2}};
\node at (1.4,3.1) {\footnotesize \ensuremath{#1}};
\end{tikzpicture}
}}}
\newcommand{\torusBlockS}[2]{\torusBlockM{#1}{#2}{S}}
\newcommand{\torusBlock}[2]{\torusBlockM{#1}{#2}{}}

\newcommand{\graphEdge}[1]{
\adjustbox{margin=0 0 0 0ex}{
\begin{tikzpicture}[scale=1]
\draw[ultra thick]  (0,0) -- (2,0);
\node at (1,-.25) {\footnotesize \ensuremath #1};
\node at (1,0.25) {\ };
\end{tikzpicture}
}}


\numberwithin{equation}{section}
\newcommand{\vecP}{\mathsf{P}}

\newcommand{\Zvir}{Z_{\rm Vir}}
\newcommand{\Ztv}{Z_{\rm CTV}}

\begin{document}
%
\renewcommand{\thefootnote}{\fnsymbol{footnote}}
\vspace{0truecm}
\thispagestyle{empty}

\begin{center}
{
\bf\LARGE
Conformal Turaev-Viro Theory
}
\end{center}


\begin{center}
{\fontsize{12}{18}\selectfont
Thomas Hartman
}

%

\vspace{.8truecm}

\centerline{{\it  Department of Physics, Cornell University, Ithaca, New York}}

\vspace{0.2cm}

 \centerline{\tt hartman@cornell.edu }

 \vspace{.25cm}

\vspace{.3cm}

\end{center}

\vspace{0.7cm}

\begin{abstract}
\noindent
We define and study Conformal Turaev-Viro (CTV) theory, a dual formulation of Virasoro TQFT based on triangulating 3-manifolds with tetrahedra. Edges of the triangulation are labeled by continuous conformal weights, and tetrahedra are glued together weighted by the Cardy density of states. We demonstrate that the CTV partition function is equal to the modular $S$-transform of the Virasoro TQFT amplitude-squared, $|\Zvir|^2$. This is analogous to a known result for discrete spin networks. The derivation uses a variant of the chain-mail formalism, adapted to the Virasoro context. As a CFT application, we derive formulae for the $S$-transforms of the squared Virasoro crossing kernels. These results lay the topological foundation to study the exact path integral of pure AdS$_3$ quantum gravity by triangulations.

\end{abstract}

\newpage

\setcounter{page}{2}
\setcounter{tocdepth}{2}
\tableofcontents
\renewcommand*{\thefootnote}{\arabic{footnote}}
\setcounter{footnote}{0}

\newpage

\section{Introduction}

Pure gravity in AdS$_3$, a quantum theory with the Einstein-Hilbert action, negative cosmological constant, and no additional degrees of freedom, is related to $SL(2,C)$ Chern-Simons theory \cite{Achucarro:1986uwr,Witten:1988hc}. The precise relationship between these two theories is very subtle --- they are not identical (see e.g. \cite{Witten:2007kt}), in part because gravity involves a sum over topologies while Chern-Simons theory is on fixed topology --- but the Chern-Simons approach, and the closely related perspective based on the quantization of Teichmuller space, can be used to calculate exact gravitational path integrals on fixed topology by the methods of topological quantum field theory (TQFT). 

There are various approaches to this topological theory. In the context of 3D gravity and the AdS/CFT correspondence, the state-of-the-art is Virasoro TQFT (VTQFT) \cite{Collier:2023fwi,Collier:2024mgv}, which is both rigorously defined and convenient for explicit calculations in gravity and TQFT. Virasoro TQFT is  believed to be equivalent to Teichmuller TQFT as formalized by Andersen and Kashaev \cite{EllegaardAndersen:2011vps}, and to the quantization of Teichmuller space developed in \cite{Verlinde:1989ua,Teschner:2001rv,Teschner:2003em,Teschner:2013tqy,Teschner:2014vca,Teschner:2014nja}. All of these approaches aim to define the quantum theory whose states are non-degenerate Virasoro conformal blocks, but the relation among them is not proven.

In this paper, we introduce a dual version of Virasoro TQFT based on triangulations. The observables of Virasoro TQFT are amplitudes $\Zvir(M_E, \Gamma(\vecP))$ where $M_E$ is a closed 3-manifold (the \textit{embedding manifold}) and $\Gamma(\vecP)$ is a framed trivalent graph $\Gamma$ labeled by conformal weights $\vecP = (P_1,\dots,P_n) \in \mathbb{R}_+^n$. In the dual approach, which we call Conformal Turaev-Viro (CTV) theory, the observables are amplitudes $\Ztv(M_E, \Gamma(\vecP))$ defined by triangulating $(M_E, \Gamma(\vecP))$ with tetrahedra, assigning a Virasoro $6j$-symbol to each tetrahedron, and integrating over the weights on internal edges. We will argue that the CTV partition function is well defined, based on various examples and consistency checks, and derive the following relationship to Virasoro TQFT:
\begin{align}\label{introFourierVV}
\Ztv(M_E, \Gamma(\vecP')) = \int_{\mathbb{R}_+^n}d\vecP (\Pi_{i=1}^n S_{P_i' P_i})
\left|
\Zvir(M_E, \Gamma(\vecP))
\right|^2
\end{align}
where $S_{P'P} = 2 \sqrt{2} \cos(4\pi P'P)$ is the Virasoro modular $S$-matrix.

We will focus on the topological theory in this paper, and discuss the applications to gravity in a separate paper \cite{gravitypaper}. But it is useful to keep those applications in mind, so let us summarize the connection to gravity briefly. The main result of \cite{gravitypaper} is that the CTV partition function is the exact gravitational path integral on a compact manifold $M$ of fixed topology,
\begin{align}\label{gravity}
Z_{A}(M, \gamma(\vecP)) = \Ztv(M_E, \Gamma(\vecP)) \ . 
\end{align}
Here $Z_A$ is the gravity amplitude on a finite spacetime region $M$, with boundary conditions that fix the dihedral angles $\psi = (\psi_1,\dots,\psi_n)$ on a set of geodesics $\gamma = (\gamma_1,\dots,\gamma_n)$ on the boundary $\p M$. The angles are related to the weights $\vecP$ by $\psi_i = 2\pi i bP_i$, with $c = 1+6(b+b^{-1})^2$ the Virasoro central charge. In the gravity context, the Fourier transform \eqref{introFourierVV} is interpreted as a transformation from fixed-length to fixed-angle boundary conditions.

The manifold $M$ where gravity lives is different from the  closed manifold $M_E$ where the graph of Virasoro TQFT or CTV theory is embedded. $M$ is related to $M_E$ by removing a regular neighborhood of the graph,  $M = M_E - N(\Gamma)$, and the cycles $\gamma$ where the boundary conditions are imposed in gravity are the meridians of the graph.

The result \eqref{gravity} is the starting point to study the exact path integral of AdS$_3$ gravity by triangulations. Unlike Virasoro TQFT, the CTV partition function \textit{manifestly} agrees with the metric formulation of gravity, making it easy to translate back and forth between semiclassical geometries and TQFT. This has many applications to the AdS$_3$/CFT$_2$ correspondence, especially in light of recent results indicating that pure gravity is dual to an ensemble of 2d CFTs \cite{Cotler:2020ugk,Chandra:2022bqq}. For details we refer to \cite{gravitypaper}.
An application to hyperbolic knot geometries and their interpretation in AdS/CFT is considered in \cite{knotspaper}.

There is a straightforward generalization of \eqref{introFourierVV} to graphs with vertices of degree $> 3$. One replaces $|\Zvir|^2$ by the relativistic (or Yokota) invariant \cite{MR1367276,MR1707988,MR1646832}, which is obtained from $|\Zvir|^2$ by integrating one or more weights with the Cardy measure. In gravity language, this corresponds to imposing fixed-angle boundary conditions on $n < 3g-3$ cycles on a genus-$g$ component of the boundary.

The identity \eqref{introFourierVV} is closely analogous to a known result for discrete spin networks \cite{Barrett:2004im}. Indeed, Conformal Turaev-Viro theory is related to Virasoro TQFT in essentially the same way that standard Turaev-Viro theory \cite{Turaev:1992hq} is related to standard TQFT. By `standard', we mean a TQFT where the spectrum is discrete and the total number of primary states (or total quantum dimension) is finite, as in most of the literature on the subject. In Virasoro/Teichmuller TQFT and CTV, the spectrum is continuous and the total number of states is infinite.

The continuous, unbounded spectrum leads to some important subtleties. The Hilbert space of Virasoro/Teichmuller TQFT is only defined on certain surfaces. In CTV, the avatar of this is that only certain triangulations are allowed. We conjecture that the theory is well defined precisely for the class of \textit{large} triangulations, defined below. Large triangulations cannot be subdivided arbitrarily, so this introduces an important difference as compared to discrete spin networks, where the triangulation can be arbitrarily fine-grained. In the context of 3D gravity, this means that CTV can be used to study macroscopic triangulations of the gravitational path integral, sliced along geodesics, but not microscopic, fine-grained triangulations. This point of view seems fundamentally different from that taken in the literature on discrete spin networks as a theory of gravity, where smooth spacetimes are supposed to emerge in the continuum limit.

Our derivation of \eqref{introFourierVV} also follows closely the literature on discrete spin networks, especially \cite{roberts1995skein,Barrett:2002vi,Garcia-Islas:2004lwa,Barrett:2004im}. We start by defining a Virasoro chain-mail invariant, based on \cite{roberts1995skein}, but with a few changes to the definitions to avoid divergences coming from the unbounded spectrum. Given a Virasoro TQFT graph $(M_E, \Gamma(\vecP))$, the chain-mail invariant is the amplitude 
\begin{align}
\Zvir(M_E, \Gamma_{CH}(\vecP)) \ , 
\end{align}
where $\Gamma_{CH}(\vecP)$ is a different graph called the chain-mail link that is constructed algorithmically from $\Gamma(\vecP)$.  To derive \eqref{introFourierVV}, we show that both sides are equal to the chain-mail amplitude by evaluating it two different ways. 

The Kashaev-Andersen Teichmuller TQFT also has a dual, Turaev-Viro-type formulation described in \cite{Kashaev:2012cz}. Just as Virasoro TQFT is believed to be equivalent to Teichmuller TQFT, it is natural to expect that the CTV theory defined in this paper is equivalent to the Turaev-Viro-type theory in \cite{Kashaev:2012cz}. The approach in \cite{Kashaev:2012cz} is based on ideal triangulations, whereas CTV is based on triangulations with ordinary, non-ideal tetrahedra. 

Another Turaev-Viro-type theory based on the Virasoro $6j$-symbol, with many of the same ingredients as CTV, was studied recently in \cite{Chen:2024unp,Hung:2024gma,Bao:2024ixc,Hung:2025vgs,Geng:2025efs}. They are not equivalent; the differences are discussed in \cite{gravitypaper}.

\subsubsection*{Examples}
Two examples that we will treat in detail are the amplitudes associated to the graphs
\begin{align}\label{introGraphs}
\cp{\footnotesize
\begin{tikzpicture}[scale=0.8]
\centerarc[very thick](0,0)(-30:90:1);
\centerarc[very thick](0,0)(90:210:1);
\centerarc[very thick](0,0)(210:335:1);
\draw[very thick] (0,0) -- (0,1);
\draw[very thick] (0,0) -- ({cos(30)},{-sin(30)});
\draw[very thick] (0,0) -- ({-cos(30)},{-sin(30)});
\node at (.9,.9) {$1$};
\node at (-.9,.9) {$2$};
\node at (-0.2,.55) {$3$};
\node at (-.55,.0) {$4$};
\node at (0.55,0) {$5$};
\node at (0,-1.25) {$6$};
\end{tikzpicture}}
\qquad \mbox{and} \qquad
\cp{\includegraphics[width=1.1in]{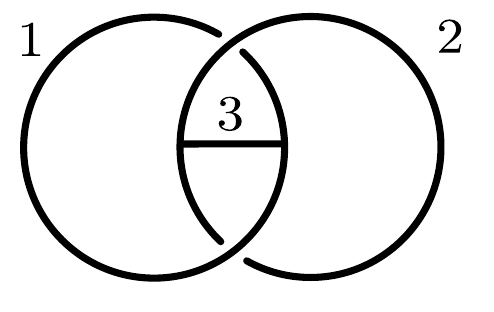}}
\end{align}
embedded in $S^3$. In the first case, the tetrahedron graph, the manifold $(S^3,\Gamma)$ can be triangulated by two tetrahedra, so the CTV partition function takes the form $(6j)^2$. The identity \eqref{introFourierVV} in this example states that the Virasoro $6j$-symbol satisfies
\begin{align}\label{introTransformJ}
\begin{Bmatrix}
P_4' & P_5' & P_6' \\
P_1' & P_2' & P_3'
\end{Bmatrix}^2
&= 
 \int_{0}^{\infty} ( \Pi_{i=1}^6 dP_i \, S_{P_i' P_i} )
 \begin{Bmatrix}
 P_1 & P_2 & P_3 \\
 P_4 & P_5 & P_6 
 \end{Bmatrix}^2 \ .
\end{align}
Thus the squared $6j$-symbol is self-dual under the Fourier transform. A similar formula holds for discrete spin networks \cite{Barrett:2002vi}.
For the second graph in \eqref{introGraphs}, the triangulation has a single tetrahedron, and the identity \eqref{introFourierVV} takes the form
\begin{align}\label{introTransformS}
\begin{Bmatrix}
P_3' & P_3' & P_1' \\
P_3' & P_3' & P_2' 
\end{Bmatrix}
&=
\int_{0}^{\infty} ( \Pi_{i=1}^3 dP_i \, S_{P_i' P_i} )
\left| \hatS_{P_1P_2}[P_3] \right|^2 \ ,
\end{align}
where $\hat{S}_{P_1P_2}[P_3]$ is the Virasoro modular $S$-matrix. 

\subsubsection*{Outline}
In section \ref{s:ctv} we define the CTV partition function, and discuss the restriction to large triangulations. In section \ref{s:vtqft} we review Virasoro TQFT, from the point of view of diagrammatic rules to calculate VTQFT amplitudes, and then derive a number of identities that form the basis of the chain-mail formalism. In section \ref{s:examples}, we show that the main identity \eqref{introFourierVV} holds in the two examples mentioned above, by explicit calculation of the CTV partition function and Fourier transforming the VTQFT amplitudes. Along the way, we derive some other interesting identities for Fourier transforms of Virasoro crossing kernels. The remainder of the paper is devoted to the derivation of \eqref{introFourierVV} for general amplitudes. In section \ref{s:chainmail} we introduce the chain-mail formalism and show that the chain-mail invariant is equal tot $\Ztv$. Then, in section \ref{s:mainidentity}, we show that its Fourier transform is $|\Zvir|^2$. Our conventions for the Virasoro crossing kernels, and some of their known properties, appear in appendix \ref{app:kernels}.

\section{Conformal Turaev-Viro Theory}\label{s:ctv}

Conformal weights in a 2d CFT are parameterized as $h = \frac{c-1}{24}+P^2$, with the Virasoro central charge $c = 1 + 6(b + b^{-1})^2$. Denote the (chiral) Cardy density of states by
\begin{align} 
\rho_0(P) &= 4\sqrt{2} \sinh(2\pi bP)\sinh(2\pi b^{-1}P) \ . 
\end{align}
Consider a closed 3-manifold $M_E$ with an embedded trivalent graph $\Gamma(\vecP)$ whose edges are labeled by weights $\vecP = (P_1,\dots, P_n) \in \mathbb{C}^n$. Choose a triangulation $T$ decomposing $M_E$ into tetrahedra, with $\Gamma \subset T$, so the  edges of the graph are part of the triangulation. Only certain types of triangulations are allowed, as we will discuss momentarily. Label the remaining edges of the triangulation by $\tilde{\vecP} = (\tilde{P}_1,\dots, \tilde{P}_m) \in \mathbb{C}^{m}$. To each tetrahedron $\Delta \in T$, we assign an amplitude given by the Virasoro $6j$-symbol,
\begin{align}\label{tetW}
W\left(
\cp{
\begin{tikzpicture}[scale=0.9]
\centerarc[very thick](0,0)(-30:90:1);
\centerarc[very thick](0,0)(90:210:1);
\centerarc[very thick](0,0)(210:335:1);
\draw[very thick] (0,0) -- (0,1);
\draw[very thick] (0,0) -- ({cos(30)},{-sin(30)});
\draw[very thick] (0,0) -- ({-cos(30)},{-sin(30)});
\node at (1,1) {$P_1$};
\node at (-1,1) {$P_2$};
\node at (-0.3,.55) {$P_3$};
\node at (-.55,.0) {$P_4$};
\node at (0.55,0) {$P_5$};
\node at (0,-1.25) {$P_6$};
\end{tikzpicture}
}
\right) 
\quad=\quad 
\begin{Bmatrix}P_4 & P_5 & P_6 \\ P_1 & P_2 & P_3 \end{Bmatrix} \ . 
\end{align}
The $6j$-symbol is the Virasoro fusion kernel up to a prefactor, as reviewed in appendix \ref{app:kernels}, and the argument of the function $W$ represents the 1-skeleton of the tetrahedron. The partition function of Conformal Turaev-Viro (CTV) theory is defined for $\vecP \in \mathbb{R}_+^n$ by the expression
\begin{align}\label{defCTV}
\Ztv(M_E, \Gamma(\vecP) )
&= \int_{\mathbb{R}_+^m} (\Pi_{i=1}^m d\tilde{P}_i \rho_0(\tilde{P}_i) )
\prod_{\Delta \in T} W(\Delta)
\end{align}
Apart from the overall normalization, this is the same formula as the standard Turaev-Viro partition function \cite{Turaev:1992hq}, with the obvious replacement of the quantum dimension by the Cardy density. In the standard Turaev-Viro theory, there is an overall factor determined by the total quantum dimension, but this diverges for Virasoro so it is omitted from the definition.

For \eqref{defCTV} to make sense, the integrals must converge and the result must be independent of the triangulation. This is not guaranteed, and without some further restrictions it is not the case. Thus one of the main problems of defining the CTV theory is to understand the class of allowed $(M_E, \Gamma, T)$, and to show that it is not empty. In this paper we will show by examples that it is not empty, and provide evidence for the following conjecture that there is a wide class of allowed triangulations.

We call the edges and vertices of the triangulation \textit{external} if they belong to $\Gamma$, and \textit{internal} otherwise. Define $V_{int}$ to be the set of internal vertices. We call a triangulation \textit{large} if $H_2(M_E - V_{int}) = 0$.

\noindent \textbf{Conjecture}. Restricted to large triangulations and graphs $\Gamma$ that are finite in Virasoro TQFT, the Conformal Turaev-Viro partition function \eqref{defCTV} is finite and independent of the triangulation.

As mentioned in the introduction, in the correspondence to pure gravity in AdS$_3$, gravity lives on the manifold  $M = M_E - N(\Gamma)$ where $N(\Gamma)$ is a regular neighborhood of the graph. $M$ is a compact manifold with boundary. Based on examples, we also conjecture that if $M$ admits a hyperbolic metric, then $(M_E,\Gamma(\vecP))$ admits a large triangulation, and therefore has a well defined CTV partition function. 

The class of large triangulations allows 2-3 and 3-2 Pachner moves, which re-triangulate two neighboring tetrahedra into three and vice-versa without adding or removing vertices. The CTV partition function is invariant under this operation by the pentagon identity. On the other hand, large triangulations do not admit 1-4 Pachner moves, which add an interior vertex, like the central vertex in the following diagram:
\begin{align}\label{badvertex}
\cp{
\begin{tikzpicture}[scale=3]
\coordinate (v3) at (0, 0.3); 
\coordinate (v2) at (0.75, 0); 
\coordinate (v1) at (1, 0.5); 
\coordinate (v4) at (0.5, 1); 
\coordinate (v5) at (0.5,0.56);
\draw[thick] (v3) -- (v2); 
\draw[thick,dashed] (v1) -- (v3); 
\filldraw[white,fill=white] (0.64,0.43) circle (0.035);
\draw[thick] (v1) -- (v5);
\filldraw[white,fill=white] (0.61,0.54) circle (0.03);
\draw[thick] (v2) -- (v5);
\draw[thick] (v3) -- (v5);
\draw[thick] (v4) -- (v5);
\draw[thick] (v3) -- (v4);
\draw[thick] (v2) -- (v1);
\draw[thick] (v2) -- (v4);
\draw[thick] (v1) -- (v4);
\end{tikzpicture}
}
\end{align}
Hence large triangulations cannot be fine-grained arbitrarily. 

The restriction to large triangulations is inspired by an identical criterion in the Andersen-Kashaev formalization of Teichmuller TQFT \cite{EllegaardAndersen:2011vps}, and by the fact that internal vertices like the one in \eqref{badvertex} produce divergences.

\section{Virasoro TQFT}\label{s:vtqft}
There is a long history of developing a topological quantum field theory associated to the quantization of Teichmuller space, whose Hilbert space is spanned by non-degenerate Virasoro conformal blocks \cite{Verlinde:1989ua,Witten:1989ip,Kashaev:1998fc,Chekhov:1999tn,Teschner:2003em,Teschner:2013tqy,Teschner:2014vca,Dimofte:2009yn}.  A rigorous definition of the theory based on ideal triangulations was given by Andersen and Kashaev \cite{EllegaardAndersen:2011vps,andersen2013new}, and their formulation is referred to as `Teichmuller TQFT'. An alternative formulation called `Virasoro TQFT' (VTQFT) was introduced in \cite{Collier:2023fwi,Collier:2024mgv}. These two theories are believed to be the same, so the name presumably refers to the starting axioms and the method of calculating, rather than to the underlying quantum theory itself. 

Virasoro/Teichmuller TQFT is not a standard TQFT, because in particular it does not have a well defined Hilbert space when cut on a 2-sphere. Nonetheless one can define and calculate its partition function on a class of 3-manifolds, including many (conjecturally all) 3-manifolds that admit a hyperbolic metric. 

We will view Virasoro TQFT as a theory that is rigorously defined by a set of diagrammatic rules to calculate amplitudes. Let $(M_E, \Gamma(\vecP))$ be a framed trivalent graph embedded in a closed 3-manifold $M_E$, with edges labeled by conformal weights $\vecP \in \mathbb{C}^n$. The framing assigns a nonzero normal vector along each edge and a cyclic ordering at each vertex. In diagrams the blackboard framing is implied, with the normal vector in the plane of projection. For admissible graphs, Virasoro TQFT assigns an amplitude written as
\begin{align}
\Zvir(M_E, \Gamma(\vecP)) \ . 
\end{align}
VTQFT amplitudes in general manifolds $M_E$ can be reduced to graphs in $S^3$ by introducing a formal weight $\Omega$, which means that the weight is integrated $\int_0^\infty dP \rho_0(P)$:
\begin{align}\label{defineOmegaLine}
\begin{overpic}[percent=true,grid=false,width=1in]{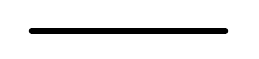}
\put (50,-5) {\footnotesize $\Omega$}
\end{overpic}
\quad=\quad
\int_0^\infty dP_1 \rho_0(P_1)
\begin{overpic}[percent=true,grid=false,width=1in]{figures/edge.pdf}
\put (50,-5) {\footnotesize $P_1$}
\end{overpic} 
\quad\equiv\quad
\int_1 
\begin{overpic}[percent=true,grid=false,width=1in]{figures/edge.pdf}
\put (50,-5) {\footnotesize $1$}
\end{overpic} \ . 
\end{align}
A label $i$ means the edge has weight $P_i$, and we use the following shorthand for integrals with the Cardy measure:
\begin{align}
\int_i = \int_0^\infty dP_i \rho_0(P_i) , \qquad
\int_{ijk \dots} = \int_i \int_j \int_k \cdots \ .
\end{align}
Insertions of $\Omega$ implement Dehn surgery on the graph, which can be used to build up any closed 3-manifold starting from $S^3$. 
Following the convention in the spin network literature, we will denote the VTQFT amplitude for a graph embedded in  $S^3$ by brackets,
\begin{align}
\langle \Gamma(\vecP) \rangle = \Zvir(S^3, \Gamma(\vecP))  \ . 
\end{align}
An edge can also be assigned the special weight $\id$, meaning the Virasoro identity representation.

\subsection{Basic rules}

The calculation of amplitudes in VTQFT can be reduced to the following set of graphical rules \cite{Collier:2023fwi,Collier:2024mgv,Post:2024itb}. Graphs with open edges should be understood as subdiagrams.

\noindent Twisting:
\begin{align}
\vbracket{
\begin{overpic}[percent=true,grid=false,width=1in]{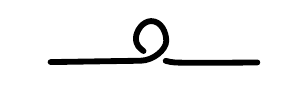}
\put (50,-5) {\footnotesize $1$}
\end{overpic}
}
=
\vbracket{\graphEdge{1}}
e^{2\pi i h_1}
\end{align}
Braiding:
\begin{align}
\vbracket{\includegraphics[width=.8in]{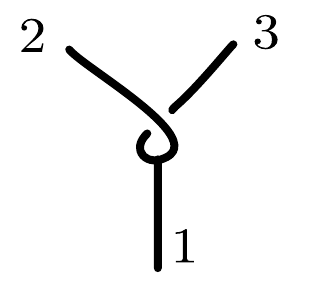}}
= \braidB_{1}^{23}
\vbracket{\includegraphics[width=.8in]{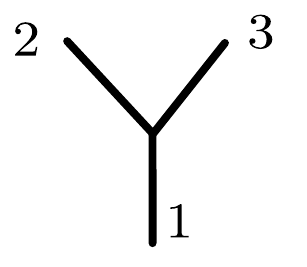}} \ , \qquad
\braidB_{1}^{23} = e^{i\pi(h_1-h_2-h_3)} \ . 
\end{align}
$\Theta$-graph:
\begin{align}
\left\langle\cp{\includegraphics[width=.75in]{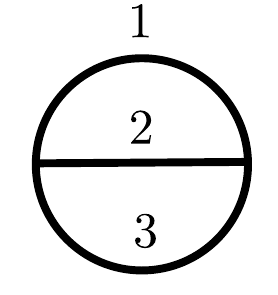}}\right\rangle
&\quad=\quad  1
\end{align}
Bubble removal:
\begin{align}\label{bubbleremoval}
\vbracket{\includegraphics[width=1.1in]{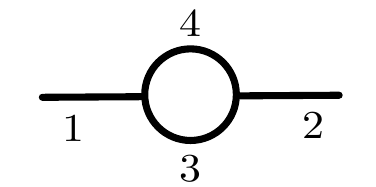}}
\quad
=
\quad
\vbracket{\graphEdge{1}}
 \quad 
\frac{1}{\rho_0(P_1)} \delta(P_1-P_2)
\end{align}
Triangle removal:
\begin{align}\label{triangleremoval}
\vbracket{
\includegraphics[width=0.9in]{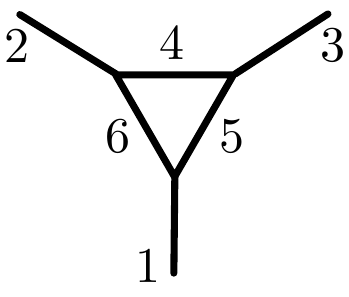}
}
\quad
=
\quad
\vbracket{\includegraphics[width=.8in]{figures/braid2.pdf}}
\begin{Bmatrix}
P_1 & P_2 & P_3 \\
P_4 & P_5 & P_6
\end{Bmatrix}
\end{align}
Fusion: 
\begin{align}
\vbracket{\graphSbig{3}{4}{2}{1}{5}}
\quad=\quad
 \int_6 
\left\langle\cp{\includegraphics[width=0.7in]{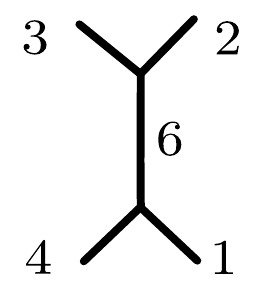}}\right\rangle
\begin{Bmatrix}P_1&P_2&P_5\\P_3&P_4&P_6\end{Bmatrix}
\end{align}
Identity fusion:
\begin{align}\label{identityfusion}
\vbracket{\includegraphics[width=.7in]{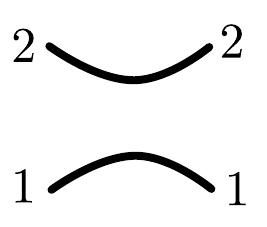}}
\quad=\quad
\int_3
\vbracket{\includegraphics[width=1in]{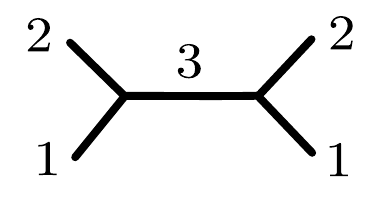}}
\end{align}
These rules give the following amplitude for a tetrahedron:
\begin{align}\label{tetamplitude}
\vbracket{\footnotesize
\begin{tikzpicture}[scale=0.8]
\centerarc[ultra thick](0,0)(-30:90:1);
\centerarc[ultra thick](0,0)(90:210:1);
\centerarc[ultra thick](0,0)(210:335:1);
\draw[ultra thick] (0,0) -- (0,1);
\draw[ultra thick] (0,0) -- ({cos(30)},{-sin(30)});
\draw[ultra thick] (0,0) -- ({-cos(30)},{-sin(30)});
\node at (.9,.9) {$1$};
\node at (-.9,.9) {$2$};
\node at (-0.2,.55) {$3$};
\node at (-.55,.0) {$4$};
\node at (0.55,0) {$5$};
\node at (0,-1.25) {$6$};
\end{tikzpicture}
}
\quad=\quad \begin{Bmatrix}P_1 & P_2 & P_3 \\ P_4 & P_5 & P_6 \end{Bmatrix}
\end{align}
Any graph can easily be reduced to the $\Theta$-graph following these rules. However, we must be careful to stay within the realm of admissible graphs. In particular we are not allowed to introduce 1-bridges (graphs that can be separated into two nontrivial components by an $S^2$ punctured by a single edge) or 0-bridges (which can be separated by an $S^2$ with no punctures). In addition, the total number of states $\int_0^\infty dP \rho_0(P)$ is infinite.
These are two essential differences from standard spin networks, so many, but  not all, of the results from ordinary TQFTs and spin networks can be carried over to Virasoro TQFT.

The amplitude for the $\Theta$-graph defines our choice of normalization of the vertices, which differs from \cite{Collier:2023fwi}. Our choice eliminates all factors of $C_0$ from VTQFT amplitudes. In gravity language, it corresponds to stripping off the factors associated to asymptotic AdS boundaries, as explained in \cite{gravitypaper}. With our normalization, there are nontrivial factors associated to inserting an identity line, which are not present in the normalization used in \cite{Collier:2023fwi}. We have
\begin{align}
\vbracket{
\begin{overpic}[percent=true,width=0.7in]{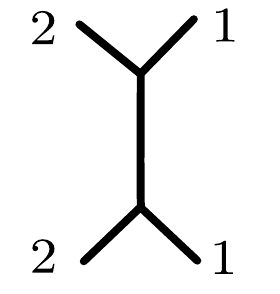}
\put (60,50) {\footnotesize $\id$}
\end{overpic}
}
\quad=\quad
\vbracket{\includegraphics[width=0.7in]{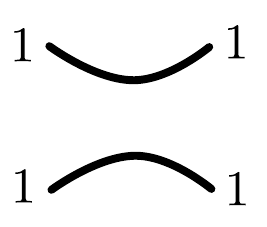}}
\quad
\frac{1}{\rho_0(P_1)}\delta(P_1-P_2)
\end{align}

\subsection{The modular $S$-matrix}
The Virasoro modular S-matrix in Racah-Wigner normalization is denoted by $\hatS_{P_1P_2}[P_3]$, following the notation in \cite{Post:2024itb}. This object is the crossing kernel for 1-point Virasoro conformal block on the torus  (up to prefactors; see appendix \ref{app:kernels}).

The crossing kernel for Virasoro characters is ($P_{1,2} \neq \id$)
\begin{align}
S_{P_1P_2} &:= \hatS_{P_1P_2}[\id] = 2\sqrt{2} \cos(4\pi P_1 P_2) \ . 
\end{align}
This satisfies the identity
\begin{align}\label{Sdelta}
\int_2 S_{P_1P_2} = \delta(P_1 - \id) 
\end{align}
where formally
\begin{align}\label{formaldelta}
\delta(P_1 -\id)= \delta(P_1 - \frac{i}{2}(b+b^{-1})) - \delta(P_1 - \frac{i}{2}(b-b^{-1})) \ .
\end{align}
The first delta function corresponds to the vacuum state, $h=0$, and the second term subtracts the null state in the vacuum representation. Although \eqref{formaldelta} is a useful heuristic, the more precise statement is that to make sense of expressions like \eqref{Sdelta} we view both sides as distributions that can be integrated against Virasoro characters, with the delta function defined to satisfy
\begin{align}
\int_0^\infty dP \chi_P(\tau) \delta(P - \id) = \chi_{\id}(\tau) \ . 
\end{align}
This is equivalent to \eqref{formaldelta} after rotating the contour of integration to the positive imaginary axis. Multiplying both sides of \eqref{Sdelta} by the character $\chi_{P_1}(\tau)$ and integrating over $P_1$, we obtain
\begin{align}
\int_0^\infty dP_1 \int_0^\infty dP_2 \rho_0(P_2) S_{P_1P_2} \chi_{P_1}(\tau) = \chi_{\id}(\tau) \ . 
\end{align}
This is a true identity, $S \cdot S \cdot \chi_{\id}(\tau) = \chi_{\id}(\tau)$, which justifies \eqref{Sdelta}.
Other expressions in Virasoro TQFT should be interpreted in the same way, as distributions that can be integrated against Virasoro conformal blocks. This means that we can use the formal delta function identity
\begin{align}\label{identitydelta}
\int_0^\infty dP_1 \delta(P_1 - \id) \langle \Gamma(P_1, P_2, \dots)\rangle
&= \langle \Gamma(\id, P_2, \dots)\rangle
\end{align}
even though the integral has no support on the identity weight.

Using the rules above, the modular $S$-matrix is a knotted handcuff graph \cite{Collier:2023fwi,deBoer:2024mqg,Post:2024itb}:
\begin{align}
\vbracket{\includegraphics[width=1.1in]{figures/handcuff.pdf}}
\quad&=\quad
\int_4 \begin{Bmatrix} P_1 &P_1 &P_3 \\ P_2 & P_2 & P_4 \end{Bmatrix}
\vbracket{\includegraphics[width=1.1in]{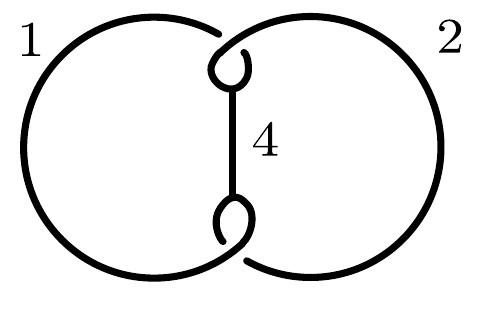}} \\
&=\quad
\int_4 \begin{Bmatrix} P_1 &P_1 &P_3 \\ P_2 & P_2 & P_4 \end{Bmatrix}
(\braidB_4^{12*})^2
\vbracket{\includegraphics[width=0.6in]{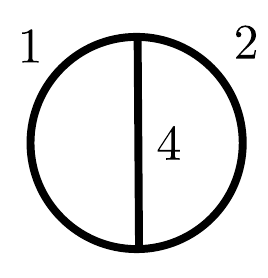}}\\
&=\quad
\int_4 \begin{Bmatrix} P_1 &P_1 &P_3 \\ P_2 & P_2 & P_4 \end{Bmatrix}
(\braidB_4^{12*})^2 \\
&= \hatS_{P_1P_2}[P_3]
\end{align}
In the last line we used eqn \eqref{SfromF}. Similar manipulations give the identities \cite{Post:2024itb}
\begin{align}\label{unlinking}
\vbracket{\includegraphics[width=0.6in]{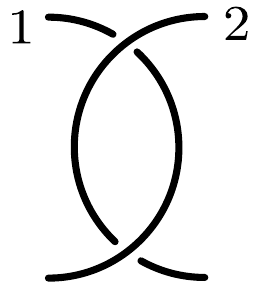}}
\quad=\quad
\int_3 \vbracket{\includegraphics[width=0.9in]{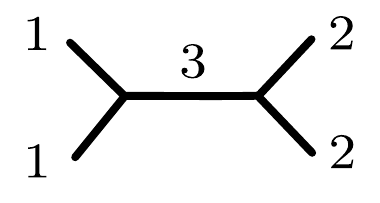}} \  \hatS_{P_1P_2}[P_3]
\end{align}
and
\begin{align}
\vbracket{\includegraphics[width=0.9in]{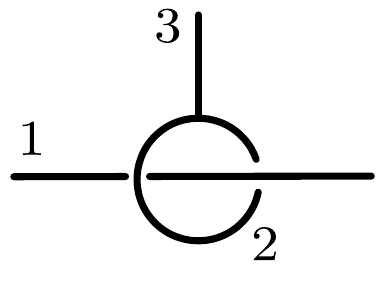}}
\quad=\quad
\vbracket{\includegraphics[width=0.8in]{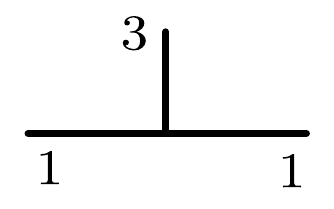}}
\quad \hatS_{P_1P_2}[P_3]
\end{align}
Setting $P_3 \to \id$ in the last formula gives
\begin{align}\label{sloop}
\vbracket{
\begin{overpic}[percent,width=1in]{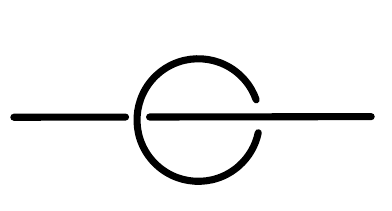}
\put (0,30) {\footnotesize 1}
\put (60,40) {\footnotesize 2}
\end{overpic}
}
\quad=\quad
\vbracket{\graphEdge{1}}\ 
\frac{ S_{P_1 P_2} }{ \rho_0(P_1)}
\end{align}
Similarly,
\begin{align}\label{fixeddoubleloop}
\vbracket{\footnotesize
\begin{overpic}[percent,width=1in]{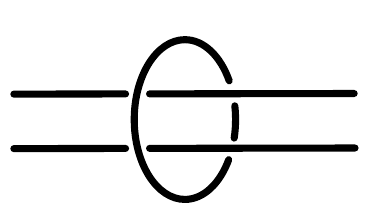}
\put (10,4) {2}
\put (10,35) {1}
\put (60,45) {3}
\end{overpic}
}
\quad=\quad
\int_0^\infty dP_4 S_{P_3 P_4} 
\vbracket{\graphSbig{1}{2}{1}{2}{4}}
\end{align}
which is derived by first fusing the two strands.

\subsection{$\Omega$-lines}
The special weight $\Omega$ was defined in \eqref{defineOmegaLine}. The effect of inserting an $\Omega$-loop is to project onto the identity inside the loop \cite{roberts1995skein}. This leads to many useful identities, and is the basis for the chain mail formalism.

\para{Identity projections}
For example,
\begin{align}
\vbracket{
\begin{overpic}[percent,width=1in]{figures/sloop.pdf}
\put (0,30) {\footnotesize 1}
\put (60,40) {\footnotesize $\Omega$}
\end{overpic}
}
\quad&=\quad \int_2 
\vbracket{
\begin{overpic}[percent,width=1in]{figures/sloop.pdf}
\put (0,30) {\footnotesize 1}
\put (60,40) {\footnotesize 2}
\end{overpic}
}\\
&=\quad \int_2 \frac{S_{P_1P_2}}{\rho_0(P_2)} \vbracket{\graphEdge{1}}\\
&=\quad
 \delta(P_1 - \id)
\vbracket{\graphEdge{1}}
%
%
\label{singleid}
\end{align}
Also
\begin{align}
\vbracket{\footnotesize
\begin{overpic}[percent,width=1in]{figures/doubleloop.pdf}
\put (10,4) {2}
\put (10,35) {1}
\put (60,45) {$\Omega$}
\end{overpic}
}
\quad=\quad
\vbracket{\includegraphics[width=0.7in]{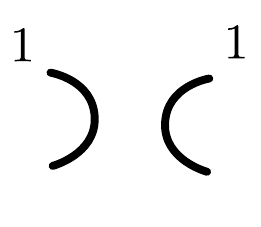}}
\ \frac{1}{\rho_0(P_1)} \delta(P_1-P_2)
\end{align}
which is derived by first fusing lines 1 and 2, then applying \eqref{singleid} and \eqref{identitydelta}. 
Looping around three edges gives the \textit{vertex join/split} identity:
\begin{align}\label{vertexjoin}
\vbracket{\footnotesize
\begin{overpic}[percent,width=1.1in]{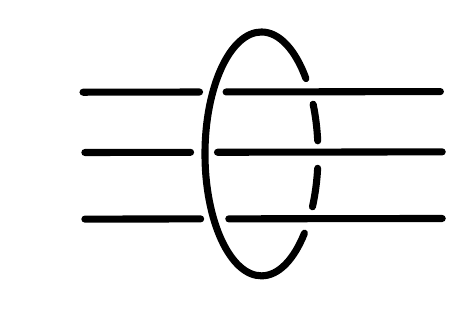}
\put (10,15) {3}
\put (10,33) {2}
\put (10, 50) {1}
\put (63, 60) {$\Omega$}
\end{overpic}
}
\quad=\quad
\vbracket{\footnotesize
\begin{overpic}[percent,width=1.1in]{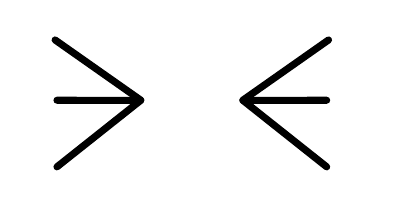}
\put (3,10) {3}
\put (3,28) {2}
\put (3, 44) {1}
\put (87,10) {3}
\put (87,28) {2}
\put (87, 44) {1}
\end{overpic}
}
\end{align}
And around four edges,
\begin{align}\label{quadvertex}
\vbracket{\footnotesize
\begin{overpic}[percent,width=1.1in]{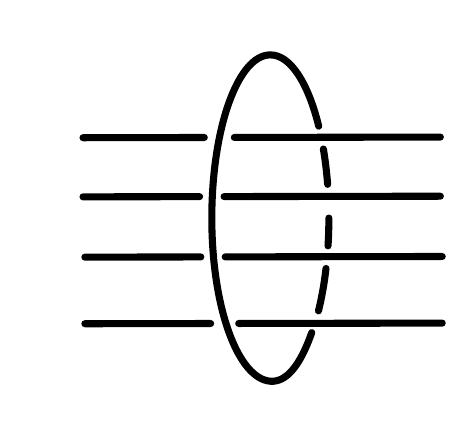}
\put (8,18) {4}
\put (8,32) {3}
\put (8,45) {2}
\put (8, 58) {1}
\put (64, 76) {$\Omega$}
\end{overpic}
}
\quad=\quad
\int_5 
\vbracket{\footnotesize
\begin{overpic}[percent,width=1.1in]{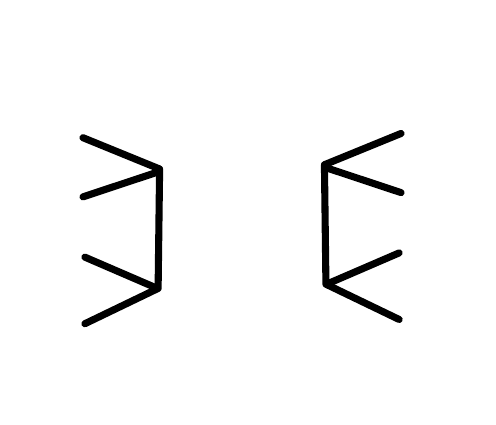}
\put (8,18) {4}
\put (8,32) {3}
\put (8,45) {2}
\put (8, 58) {1}
\put (35, 38) {5}
\put (55, 38) {5}
\put (87,18) {4}
\put (87,32) {3}
\put (87,45) {2}
\put (87, 58) {1}
\end{overpic}
}
\end{align}

\para{Intermediate state projection}
By linking $\Omega$ with another loop, we can also project onto non-vacuum intermediate states. For example,
\begin{align}\label{doublelooplink}
\vbracket{\footnotesize
\begin{overpic}[percent,width=1.2in]{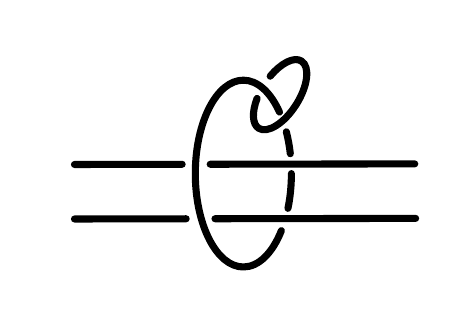}
\put (7,15) {2}
\put (7,33) {1}
\put (70, 55) {3}
\put (33,46) {$\Omega$}
\end{overpic}
}
&\quad=\quad \int_0^\infty dP_4 S_{P_3P_4}
\vbracket{\footnotesize
\begin{overpic}[percent,width=1in]{figures/doubleloop.pdf}
\put (10,4) {2}
\put (10,35) {1}
\put (60,45) {4}
\end{overpic}
}\\
&\quad=\quad
\int_0^{\infty}dP_4 S_{P_3P_4} \int_0^\infty dP_5  S_{P_4 P_5}
\vbracket{\graphSbig{1}{2}{1}{2}{5}}\\
&\quad=\quad \int_0^\infty dP_5 \delta(P_5-P_3) 
\vbracket{\graphSbig{1}{2}{1}{2}{5}}\\
&\quad=\quad 
\vbracket{\graphSbig{1}{2}{1}{2}{3}}
\end{align}

\para{Fourier transforms}
To take the Fourier transform of a VTQFT amplitude, assuming each edge has an independent label, we can simply replace each edge 
\begin{align}
\cp{\graphEdge{1}}
\quad \longrightarrow \quad
\cp{\begin{overpic}[percent,width=1.1in]{figures/sloop.pdf}
\put (0,30) {\footnotesize $\Omega$}
\put (60,40) {\footnotesize $1'$}
\end{overpic}}
\quad=\quad
\int_0^\infty dP_1 S_{P_1' P_1} 
\cp{\graphEdge{1}}
\end{align}

\para{Handleslides}
By a sequence of two identity fusions, one finds that $\Omega$'s can freely slide around each other:
\begin{align}
\vbracket{
\footnotesize
\begin{overpic}[percent,width=1in]{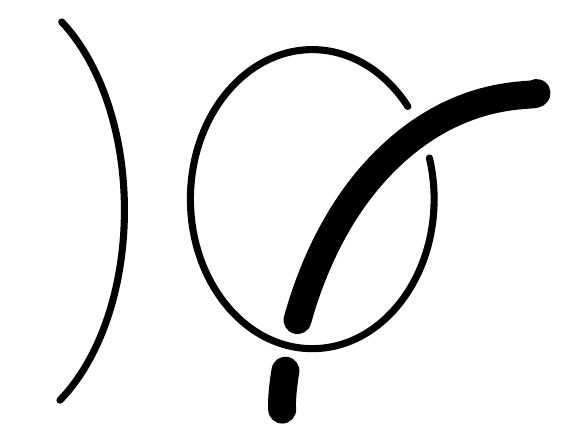}
\put (5,50) {$\Omega$}
\put (40,40) {$\Omega$}
\end{overpic}
}
\quad=\quad
\vbracket{
\footnotesize
\begin{overpic}[percent,width=1in]{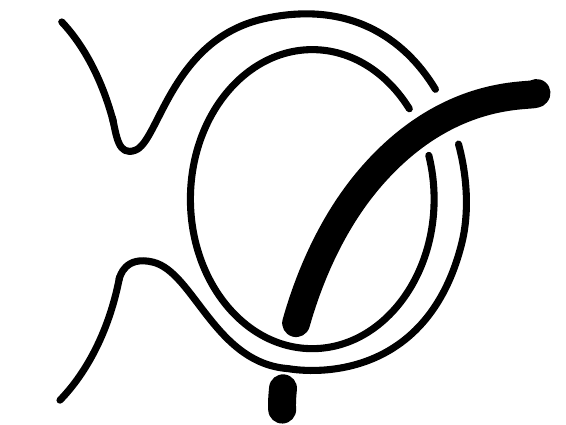}
\put (5,50) {$\Omega$}
\put (40,40) {$\Omega$}
\end{overpic}
}
\end{align}
where the thick line represents any subgraph.

\section{Examples}\label{s:examples}
Before giving the general derivation, we will check the main identity
\begin{align}\label{mainidex}
\Ztv(M_E, \Gamma(\vecP')) = \int_{\mathbb{R}_+^n}d\vecP (\Pi_{i=1}^n S_{P_i' P_i})
\left|
\Zvir(M_E, \Gamma(\vecP))
\right|^2
\end{align}
in two examples: the tetrahedron graph and the knotted handcuff, embedded in $S^3$. In each case we calculate the Fourier transform of $|\Zvir|^2$ and check that it agrees with the triangulation by direct comparison, leaving the explanation to later sections. Along the way we will find some interesting identities for the $S$-transforms of the Virasoro crossing kernels.

\subsection{$6j$-symbol}
The first example is the tetrahedron graph, 
\begin{align}\label{gammaTet}
\Gamma(\vecP) &\quad=\quad \cp{\footnotesize
\begin{tikzpicture}[scale=0.8]
\centerarc[ultra thick](0,0)(-30:90:1);
\centerarc[ultra thick](0,0)(90:210:1);
\centerarc[ultra thick](0,0)(210:335:1);
\draw[ultra thick] (0,0) -- (0,1);
\draw[ultra thick] (0,0) -- ({cos(30)},{-sin(30)});
\draw[ultra thick] (0,0) -- ({-cos(30)},{-sin(30)});
\node at (.9,.9) {$1$};
\node at (-.9,.9) {$2$};
\node at (-0.2,.55) {$3$};
\node at (-.55,.0) {$4$};
\node at (0.55,0) {$5$};
\node at (0,-1.25) {$6$};
\end{tikzpicture}}
\end{align}
with amplitude
\begin{align}
\langle \Gamma(\vecP) \rangle = \begin{Bmatrix}P_1 & P_2 & P_3 \\ P_4 & P_5 & P_6 \end{Bmatrix} \ . 
\end{align}
\subsubsection{Fourier transforms}
Let us calculate the Fourier transform of $|\langle \Gamma(\vecP)\rangle|^2$ on all six entries. This is similar to a calculation for discrete spin networks in \cite{Barrett:2002vi}, with a few small changes to avoid diagrams that are undefined in VTQFT. Consider the amplitude
\begin{align}
I(\vecP)
\quad=\quad
\vbracket{\footnotesize \begin{overpic}[percent,width=1.5in]{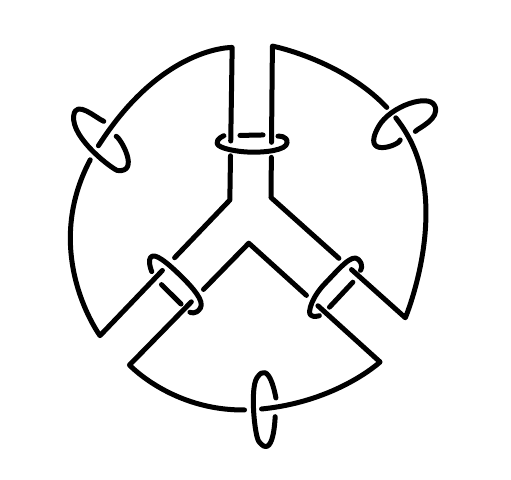}
\put (85,75) {1}
\put (10,74) {2}
\put (57,64) {3}
\put (54,0) {6}
\put (25,45) {4}
\put (71,45) {5}
\put (7,40) {$\Omega$}
\put (85,40) {$\Omega$}
\put (30,9) {$\Omega$}
\end{overpic}}
\end{align}
Using the vertex join identity \eqref{vertexjoin} on all three $\Omega$-loops gives
\begin{align}\label{sixIA}
I(\vecP) \quad=\quad 
\vbracket{\footnotesize \begin{overpic}[percent,width=1in]{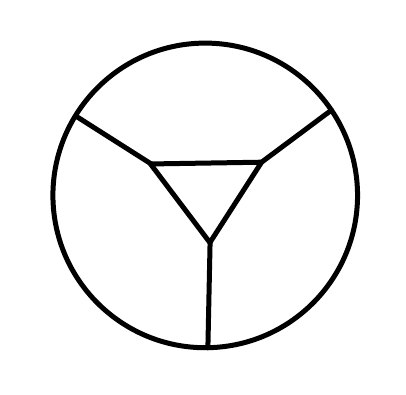}
\put (75,55) {1}
\put (23,53) {2}
\put (49,60) {3}
\put (38,38) {4}
\put (60,38) {5}
\put (53, 20) {6}
\put (49,90) {3}
\put (15,15) {4}
\put (83,15) {5}
\end{overpic}
}
\quad=\quad \begin{Bmatrix}P_4 & P_5 & P_6 \\ P_1 & P_2 & P_3 \end{Bmatrix}^2 \ . 
\end{align}
Alternatively, we can evaluate $I(\vecP)$ as follows:
\begin{align}
I(\vecP) 
&\quad=\quad
\int_{\mathbb{R}_+^3} dP_1' dP_2' dP_6' S_{P_1P_1'}S_{P_2P_2'}S_{P_6P_6'} 
\vbracket{\footnotesize\begin{overpic}[percent,width=1.3in]{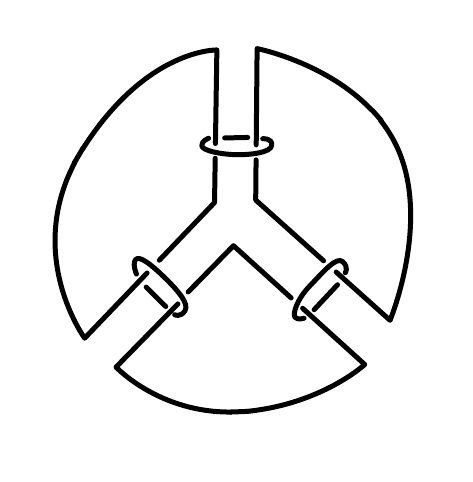}
\put (15,75) {$2'$}
\put (82,76) {$1'$}
\put (48,6) {$6'$}
\put (59,67) {3}
\put (22,45) {4}
\put (73,45) {5}
\end{overpic}
}\\
&\quad=\quad
\int_{\mathbb{R}_+^6}(\Pi_{i=1}^6 dP_i' S_{P_i P_i'}) 
\vbracket{\footnotesize\begin{overpic}[percent,width=1in]{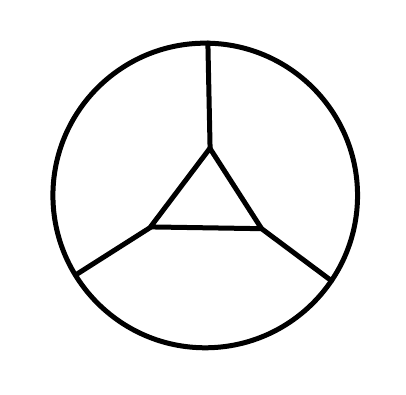}
\put (83,72) {$1'$}
\put (12,72) {$2'$}
\put (48,-2) {$6'$}
\put (53,70) {$3'$}
\put (27,25) {$4'$}
\put (65,25) {$5'$}
\put (61,51) {$1'$}
\put (36,50) {$2'$}
\put (48,30) {$6'$}
\end{overpic}}
\label{trisq}
\end{align}
where we used \eqref{sloop} in the first line and \eqref{fixeddoubleloop} in the second line. The amplitude in the integrand on the last line is again $(6j)^2$, but with the rows swapped as compared to \eqref{sixIA}. Therefore, the Virasoro $6j$-symbol satisfies the self-duality relation
\begin{align}\label{TransformJ}
\begin{Bmatrix}
P_4' & P_5' & P_6' \\
P_1' & P_2' & P_3'
\end{Bmatrix}^2
&= 
 \int_{0}^{\infty} ( \Pi_{i=1}^6 dP_i \, S_{P_i' P_i} )
 \begin{Bmatrix}
 P_1 & P_2 & P_3 \\
 P_4 & P_5 & P_6 
 \end{Bmatrix}^2 \ .
\end{align}
We can also do an $S$-transform on a subset of the arguments. This is not relevant to our main discussion, but we note one nice relation for the Fourier transform of just the top row:
\begin{align}\label{sixjRowTransform}
\int_0^\infty &( \Pi_{i=1}^3 dP_i \, S_{P_i' P_i} ) 
 \begin{Bmatrix}
 P_1 & P_2 & P_3 \\
 P_4 & P_5 & P_6 
 \end{Bmatrix}^2 \\
 &= \int_0^\infty dP_7 \rho_0(P_7) 
 \hatS_{P_1'P_6}^*[P_7]
 \hatS_{P_6P_2'}[P_7]
 \hatS_{P_2'P_4}^*[P_7]
 \hatS_{P_4P_3'}[P_7]
 \hatS_{P_3'P_5}^*[P_7]
 \hatS_{P_5P_1'}[P_7] \ .
 \notag 
\end{align}
This is derived by starting with $ \begin{Bsmallmatrix}
 P_1 & P_2 & P_3 \\
 P_4 & P_5 & P_6 
 \end{Bsmallmatrix}^2$ in the form of the bracketed graph in the integrand of \eqref{trisq}, adding loops around 1,2, and 3 to implement the Fourier transform, and then integrating over 1,2, and 3 using \eqref{identityfusion}. This produces the diagram
 \begin{align}
\vbracket{\footnotesize\begin{overpic}[percent,width=1.3in]{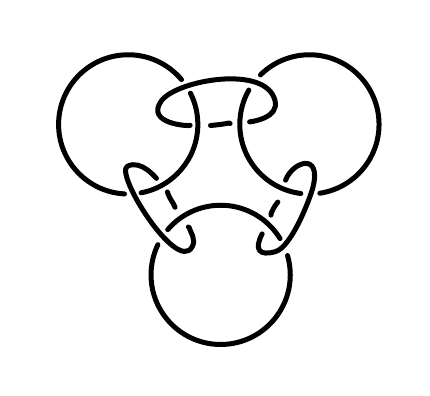}
\put (86,75) {6}
\put (12,75) {5}
\put (50,3) {4}
\put (50,76) {$1'$}
\put (72,37) {$2'$}
\put (25,36) {$3'$}
\end{overpic}}
\end{align}
which, using \eqref{unlinking} several times, evaluates to the right-hand side of \eqref{sixjRowTransform}. The transform \eqref{sixjRowTransform} can also be derived from the pentagon identity in the form $(6j)^2 = \int (6j)(6j)(6j)$ by applying the Post-Tsiares formula \cite{Post:2024itb} three times.

\subsubsection{Triangulation}
Now we turn to the CTV partition function, $\Ztv(S^3, \Gamma(P))$, with $\Gamma(\vecP)$ the tetrahedral graph in \eqref{gammaTet}. The first step is to triangulate $S^3$ with tetrahedra, such that the embedded graph $\Gamma(\vecP) \subset S^3$ is part of the triangulation. This can be done with two tetrahedra, identical up to an orientation-reversal:
\begin{align}
\cp{
\begin{tikzpicture}[scale=3]
\coordinate (v3) at (0, 0.3); 
\coordinate (v2) at (0.75, 0); 
\coordinate (v1) at (1, 0.5); 
\coordinate (v4) at (0.5, 1); 
\draw[thick] (v3) -- (v2) node[midway, below] {$P_6$};
\draw[thick,dashed] (v1) -- (v3) node[pos=0.65,above] {$P_5$};
\filldraw[white,fill=white] (0.64,0.43) circle (0.035);
\draw[thick] (v3) -- (v4) node[midway,left] {$P_1$};
\draw[thick] (v2) -- (v1) node[midway,right] {$P_4$};
\draw[thick] (v2) -- (v4) node[pos=0.7,left] {$P_2$};
\draw[thick] (v1) -- (v4) node[midway,right] {$P_3$};
\end{tikzpicture}
\qquad \qquad
\begin{tikzpicture}[scale=3]
\coordinate (v3) at (0, 0.3); 
\coordinate (v2) at (0.75, 0); 
\coordinate (v1) at (1, 0.5); 
\coordinate (v4) at (0.5, 1); 
\draw[thick] (v3) -- (v2) node[midway, below] {$P_4$};
\draw[thick,dashed] (v1) -- (v3) node[pos=0.65,above] {$P_5$};
\filldraw[white,fill=white] (0.64,0.43) circle (0.035);
\draw[thick] (v3) -- (v4) node[midway,left] {$P_3$};
\draw[thick] (v2) -- (v1) node[midway,right] {$P_6$};
\draw[thick] (v2) -- (v4) node[pos=0.7,left] {$P_2$};
\draw[thick] (v1) -- (v4) node[midway,right] {$P_1$};
\end{tikzpicture}
}
\end{align}
The tetrahedra are glued together on their faces, and the graph $\Gamma(\vecP)$ lives on the edges. There are no internal edges in this example. From the definitions \eqref{tetW}-\eqref{defCTV} we find the CTV partition function
\begin{align}
\Ztv(S^3, \Gamma(\vecP)) &= \begin{Bmatrix}
P_4 & P_5 & P_6\\
P_1 & P_2 & P_3
\end{Bmatrix}^2 \ .
\end{align}
This is indeed the Fourier transform of $|\langle \Gamma(\vecP)\rangle|^2$ as shown in \eqref{TransformJ}. Thus we have confirmed the main identity \eqref{mainidex} in this example.

\subsection{Modular S-matrix}
In the next example, we consider the knotted handcuff graph
\begin{align}\label{gammahandcuff}
\Gamma(\vecP) &\quad=\quad \vbracket{\includegraphics[width=1.1in]{figures/handcuff.pdf}}
\end{align}
with amplitude
\begin{align}
\langle \Gamma(\vecP)\rangle = \hatS_{P_1P_2}[P_3] \ . 
\end{align}
\subsubsection{Fourier transforms}
We start by representing $|\langle \Gamma(\vecP)\rangle|^2$ as a single graph,
\begin{align}\label{ssquareddiagram}
|\hatS_{P_1P_2}[P_3]|^2
&\quad=\quad 
\vbracket{\footnotesize \begin{overpic}[percent,width=1.2in]{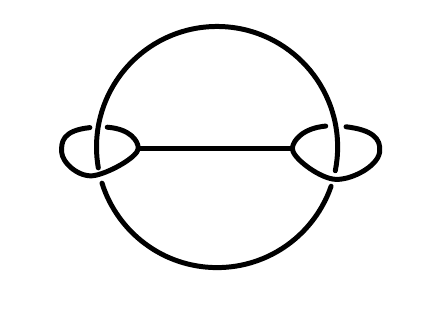}
\put (8, 38) {1}
\put (90,38) {1}
\put (48,43) {3}
\put (48,15) {2}
\end{overpic}}
\end{align}
This is derived by unlinking the two 1-loops using \eqref{unlinking}, then removing the bubbles with \eqref{bubbleremoval} \cite{Post:2024itb}. The Fourier transform with respect to $P_2$ is
\begin{align}\label{sft1a}
\int_0^\infty dP_2 S_{P_2'P_2} | \hatS_{P_1P_2}[P_3] |^2
&\quad= \quad
\vbracket{\footnotesize \begin{overpic}[percent,width=1.2in]{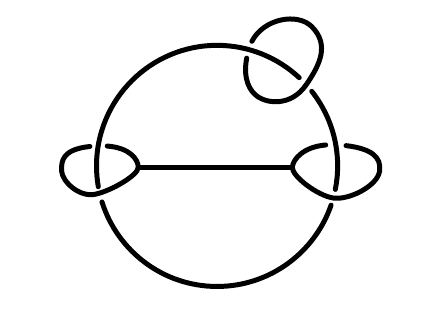}
\put (8, 30) {1}
\put (90,30) {1}
\put (48,36) {3}
\put (48,10) {$\Omega$}
\put (78, 67) {$2'$}
\end{overpic}}\\
&\quad=\quad
\vbracket{\footnotesize
\begin{tikzpicture}[scale=0.8]
\centerarc[very thick](0,0)(-30:90:1);
\centerarc[very thick](0,0)(90:210:1);
\centerarc[very thick](0,0)(210:335:1);
\draw[very thick] (0,0) -- (0,1);
\draw[very thick] (0,0) -- ({cos(30)},{-sin(30)});
\draw[very thick] (0,0) -- ({-cos(30)},{-sin(30)});
\node at (.9,.9) {$1$};
\node at (-.9,.9) {$1$};
\node at (-0.2,.55) {$2'$};
\node at (-.55,.0) {$1$};
\node at (0.55,0) {$1$};
\node at (0,-1.25) {$3$};
\end{tikzpicture}
}\label{sft1i}\\
&= \quad \begin{Bmatrix} P_1 & P_1 & P_2' \\ P_1 & P_1 & P_3 \end{Bmatrix}
\label{sft1}
\end{align}
In the second line we used \eqref{vertexjoin}. The identity  \eqref{sft1} is a special case of the Post-Tsiares formula \cite{Post:2024itb}, reproduced in \eqref{ptformula}. Next we transform with respect to $P_3$. Starting with the diagram in \eqref{sft1i} gives
\begin{align}
\int_0^\infty dP_2 dP_3 S_{P_2'P_2}S_{P_3'P_3} | \hatS_{P_1P_2}[P_3] |^2
&= \vbracket{\footnotesize \begin{overpic}[percent,width=1.2in]{figures/ssquared.pdf}
\put (8, 38) {1}
\put (90,38) {1}
\put (48,43) {$2'$}
\put (48,15) {$3'$}
\end{overpic}}
\\
&= |\hatS_{P_1 P_3'}[P_2'] |^2
\end{align}
where we used \eqref{fixeddoubleloop} in the first line and \eqref{ssquareddiagram} in the second.  The last Fourier transform, with respect to $P_1$, is the same calculation as \eqref{sft1a}-\eqref{sft1}. Thus the final result for the $S$-transform is
\begin{align}\label{TransformS}
\begin{Bmatrix}
P_3' & P_3' & P_1' \\
P_3' & P_3' & P_2' 
\end{Bmatrix}
&=
\int_{0}^{\infty} ( \Pi_{i=1}^3 dP_i \, S_{P_i' P_i} )
\left| \hatS_{P_1P_2}[P_3] \right|^2 \ .
\end{align}

\subsubsection{Triangulation}\label{sss:ctvSmatrix}
Let us compare to the CTV partition function on $(S^3, \Gamma(\vecP))$, with $\Gamma(\vecP)$ the knotted handcuff in \eqref{gammahandcuff}. This manifold can be triangulated by a single tetrahedron, 
\begin{align}
\cp{
\begin{tikzpicture}[scale=3]
\coordinate (v3) at (0, 0.3); 
\coordinate (v2) at (0.75, 0); 
\coordinate (v1) at (1, 0.5); 
\coordinate (v4) at (0.5, 1); 
\draw[thick] (v3) -- (v2) node[midway, below] {$P_3$};
\draw[thick,dashed] (v1) -- (v3) node[pos=0.65,above] {$P_2$};
\filldraw[white,fill=white] (0.64,0.43) circle (0.035);
\draw[thick] (v3) -- (v4) node[midway,left] {$P_3$};
\draw[thick] (v2) -- (v1) node[midway,right] {$P_3$};
\draw[thick] (v2) -- (v4) node[pos=0.7,left] {$P_1$};
\draw[thick] (v1) -- (v4) node[midway,right] {$P_3$};
\node[left] at (v3) {$v_3$};
\node[below] at (v2) {$v_2$};
\node[right] at (v1) {$v_1$};
\node[above] at (v4) {$v_4$};
\end{tikzpicture}
}
\end{align}
where we have labeled the vertices $v_1,\dots,v_4$. The two front faces of the tetrahedron are glued together, and the two back faces are glued together, so the vertices are identified as $v_3 \equiv v_1$ and $v_2 \equiv v_4$. For example, the faces $(v_1v_3v_4)$ and $(v_1v_3v_2)$ of the tetrahedron are the front and back of the surface shaded here on the embedded graph:
\begin{align}
\cp{\includegraphics[width=1.3in]{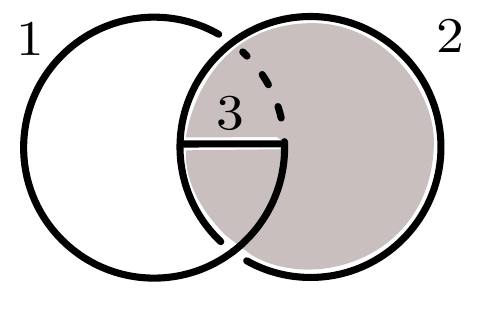}}
\end{align}
Since the triangulation has just one tetrahedron, and no internal edges, the CTV partition function is simply
\begin{align}
\Ztv(S^3, \Gamma(\vecP)) &= \begin{Bmatrix} P_3 & P_3 & P_1 \\ P_3 & P_3 & P_2\end{Bmatrix} \ . 
\end{align}
This is the $S$-transform of $| \langle \Gamma(\vecP) \rangle|^2 = |\hatS_{P_1P_2}[P_3]|^2$ as demonstrated in the previous subsection, confirming the main identity \eqref{mainidex} in this example.

\section{Virasoro Chain Mail} \label{s:chainmail}

The chain mail formalism was introduced by Roberts \cite{roberts1995skein} to prove the Turaev-Viro conjecture \cite{Turaev:1992hq}  for discrete spin networks, which states that $Z_{TV}(M) = |\langle \Gamma \rangle|^2$ without any fixed, external edges. The conjecture had been proved earlier using the shadow formalism \cite{Turaev:1994xb}, but the chain mail method is easier and more intuitive. Barrett, Garcia-Islas, and Martins \cite{Barrett:2004im} later generalized this approach to allow for external edges. In what follows we will extend \cite{Barrett:2004im} to the Virasoro case. The logic follows \cite{roberts1995skein,Barrett:2004im} very closely, so this is mostly review, but some of the details of the derivation are different, because the original derivations use graphs that are disallowed in Virasoro TQFT. Our task is to rephrase the arguments in a way that avoids disallowed graphs, which is not difficult.

We will give two different descriptions of the chain-mail invariant: One starting from a VTQFT graph, and the second starting from a Heegaard splitting of a 3-manifold. Then we show that the two descriptions are equivalent, and that the chain-mail invariant is equal to the CTV partition function:
\begin{align}\label{ctvcm}
\Ztv(M_E, \Gamma(\vecP)) = \Zvir(M_E, \Gamma_{CH}(\vecP)) \ . 
\end{align}

For the next few sections, we will focus on VTQFT graphs embedded in $S^3$, with $\vecP \in \mathbb{C}^n$.The results are extended to graphs in other 3-manifolds $M_E$ by adding surgery links in section  \ref{ss:generalME}.

\subsection{The chain-mail link}\label{ss:chainmailGraph}
Consider a VTQFT graph $(S^3, \Gamma(\vecP))$, with amplitude $\langle \Gamma(\vecP)\rangle$. The corresponding \textit{chain-mail link} $\Gamma_{CH}(\vecP)$ is defined by the following procedure. Draw the projection of $\Gamma$ in the plane with over- and under-crossings, and replace
\begin{align}\label{cmrules}
\cp{\footnotesize \graphEdge{i}}
\quad &\longrightarrow \quad
\cp{\footnotesize
\begin{overpic}[percent,width=1in]{figures/doubleloop.pdf}
\put (60,45) {$i$}
\end{overpic}\notag
}\\
\cp{\includegraphics[width=1.2in]{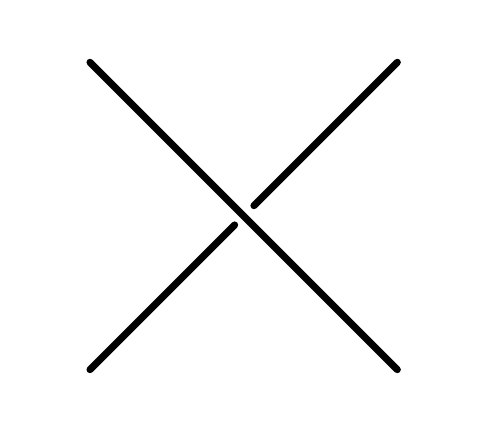}}
\quad &\longrightarrow \quad
\cp{\includegraphics[width=1.2in]{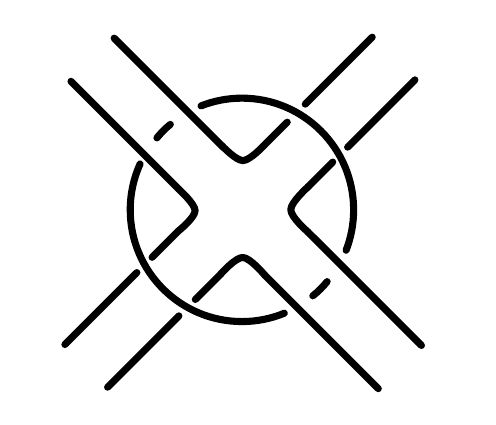}}\\
\cp{\includegraphics[width=0.75in]{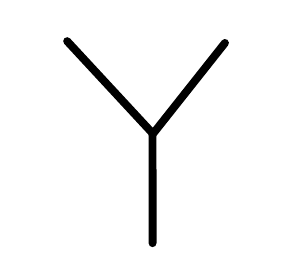}}
\quad &\longrightarrow \quad
\cp{\includegraphics[width=0.75in]{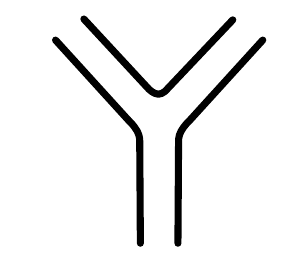}}\notag
\end{align}
%
Then delete the outermost loop that forms the perimeter of the original graph, and assign weight $\Omega$ to all unlabeled edges.\footnote{This is the same procedure as in \cite{Barrett:2004im} up to one change: in \cite{Barrett:2004im} the perimeter loop is not deleted. This loop can be unlinked from the rest of the graph by handleslides, so it produces an overall factor of the total quantum dimension which only affects the overall normalization for discrete spin networks but diverges for Virasoro.} The resulting graph is $\Gamma_{CH}(\vecP)$, and the  \textit{Virasoro chain-mail invariant} is the VTQFT amplitude for this graph,
\begin{align}
\Zvir(S^3, \Gamma_{CH}(\vecP)) \ . 
\end{align}
For example, let's apply this procedure to the graph
\begin{align}\label{gammahandcuffB}
\Gamma(\vecP) \quad=\quad \cp{\includegraphics[width=1.3in]{figures/handcuff.pdf}}
\end{align}
The result is the corresponding chain-mail link,
\begin{align}\label{gammaCHhandcuff}
\Gamma_{CH}(\vecP) \quad= \quad
\cp{\footnotesize \begin{overpic}[width=2.5in,percent]{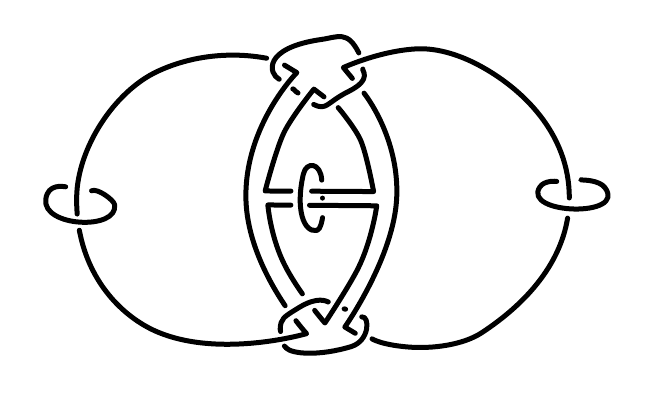}
\put (2,30) {1}
\put (97,30) {2}
\put (49,37) {3}
\end{overpic}}
\end{align}
In this graph and from now on, the unlabeled loops implicitly have weight $\Omega$.

\subsection{Marked Heegaard Diagrams}\label{ss:heegaard}

We have defined the chain mail link by a graphical procedure. Now we will describe it geometrically in terms of a marked Heegaard diagram.\footnote{This construction corrects what appears to be a mistake or typo in \cite{Barrett:2004im}. For example, the definition of a generalized Heegaard diagram there only allows up to two independent $P_i$'s in a genus-2 splitting, when there should be three. This change does not affect their conclusions.} Background can be found in the review article \cite{scharlemann2000heegaard}.
 
First let us establish some terminology. A compression body is a 3-manifold $M$ constructed by starting with a genus-$g$ Riemann surface $\Sigma_+$, thickening it to $\Sigma_+ \times \mbox{Interval}$, and then filling in $k \leq g$ independent cycles $\tilde{\gamma} = (\tilde{\gamma}_1,\dots,\tilde{\gamma}_k)$.   If $k=g$, then this reduces to a handlebody with $\p M = \Sigma_+$, while for $k < g$ there is an additional boundary component $\Sigma_-$, with $\p M = \Sigma_- \sqcup \Sigma_+$. 

Define a \textit{marked compression body} to be a compression body together with a choice of pants decomposition on $\Sigma_-$. Suppose $\Sigma_-$ has connected components $\Sigma_-^a$ for $a=1,2,\dots$. Then the marking consists of $3g_a-3$ separating curves on each component of genus $g_a\geq 2$, and one simple closed curve on each torus component.\footnote{In other words, we choose an OPE channel on each component of $\Sigma_-$.} The marking can be described as a set of disjoint simple closed curves $\gamma = (\gamma_1, \dots, \gamma_n) \subset \Sigma_+$. Therefore, a marked compression body is specified by $(\Sigma_+, \tilde{\gamma}, \gamma)$, a Riemann surface $\Sigma_+$ and two sets of closed curves on $\Sigma_+$. For example, the diagram
\begin{align}\label{markedCB}
\cp{\begin{overpic}[percent,width=4in]{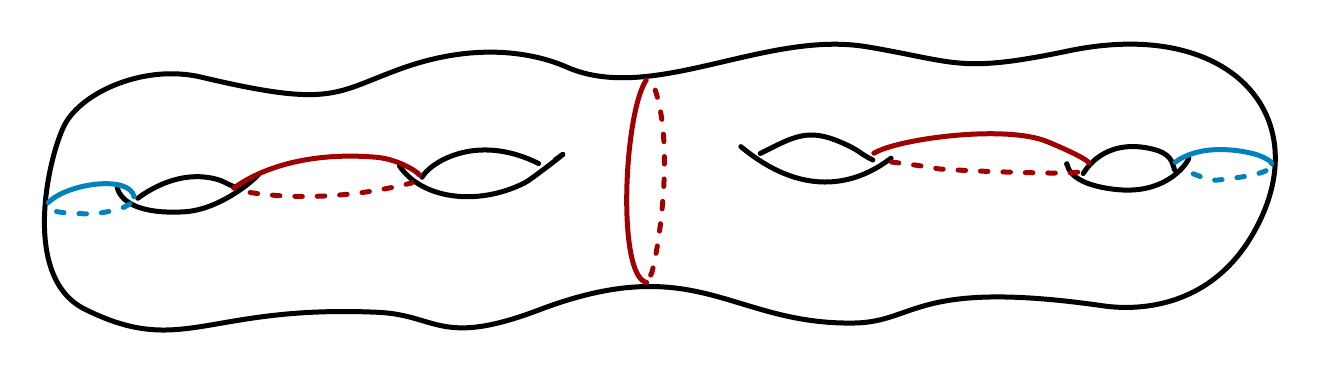}
\put (6,9) {$\tilde{\gamma}_1$}
\put (23,11) {$\gamma_1$}
\put (43,12) {$\gamma_2$}
\put (73,13) {$\gamma_3$}
\put (90,12) {$\tilde{\gamma}_2$}
\end{overpic}
}
\end{align}
represents a compression body where $\Sigma_+$ has genus 4, and the two blue cycles are filled. The inner boundary $\Sigma_-$ is a Riemann surface of genus 2, separated by the red curves.

Given a 3-manifold $M$ and any bipartition of its boundary $\p M = \Sigma_1 \sqcup \Sigma_2$, one can find a Heegaard splitting $(C_1, C_2, \Sigma)$ that decomposes $M$ as the union of two compression bodies $C_1$ and $C_2$. The compression bodies are glued on their $+$ boundaries, $\Sigma = \Sigma_+(C_1) = \Sigma_+(C_2)$, and have $\Sigma_-(C_1) = \Sigma_1$, $\Sigma_-(C_2) = \Sigma_2$. 

Let us choose $\Sigma_2 = \emptyset$. This results in a Heegaard splitting $(C, H, \Sigma)$ into a compression body $C$ and a handlebody $H$, with $\p C = \p M$. Supplying a marking on $C$ gives a marked Heegard splitting of $M$. A \textit{marked Heegard diagram} $(\Sigma, \gamma, \tilde{\gamma}, \rho)$ consists of the splitting surface $\Sigma$, the marking $\gamma$, the cycles $\tilde{\gamma}$ that contract inside the compression body, and the cycles $\rho = (\rho_1,\dots,\rho_g)$ that contract inside the handlebody. Any 3-manifold with a marked boundary can be characterized by a marked Heegaard diagram. For example, \eqref{markedCB} can be turned into a marked Heegaard diagram by adding four independent cycles $\rho$ to fill in on the handlebody side of $M$.

Sometimes we label the unfilled $\gamma$ cycles by $\vecP = (P_1,\dots,P_n)$, and we will also refer to $(\Sigma, \gamma(\vecP), \tilde{\gamma}, \rho)$ as a marked Heegaard diagram.

As an aside, let us note that a marked Heegaard diagram is exactly the data that needs to be supplied as the argument of the gravitational path integral on a fixed-topology compact 3-manifold with boundary. The topology of the 3-manifold $M$ is specified by the unmarked Heegaard diagram, and the boundary conditions on $\p M$ are specified by the marking.

\subsection{Chain mail from a Heegaard splitting}\label{ss:cmheegaard}
Any 3-manifold can be triangulated, and a triangulation also provides a Heegaard splitting. Consider first the case that $M$ is closed. Starting from a triangulation $T$, thicken the edges of the triangulation to a regular neighborhood $N(T)$, which is a neighborhood of $T$ that deformation retracts to $T$. $N(T)$ is a handlebody $H_1$, and the complement $M - N(T)$ is the second handlebody, $H_2$, so this defines a Heegaard splitting on the surface $\p N(T)$. 

Now we are ready to describe the Heegaard construction of the chain mail invariant. Let $(M_E, \Gamma(\vecP))$ be a VTQFT graph embedded in a closed 3-manifold $M_E$. We assume that $(M_E, \Gamma)$ admits a large triangulation $T$, as defined below \eqref{defCTV}, with $\Gamma \subset T$. (Formally, the following argument does not require the triangulation to be large, but we will see below that a small triangulation leads to a divergent chain mail invariant.) Partition the triangulation into $T = \Gamma \cup \tilde{\Gamma}$, with external edges $\Gamma$ and internal edges $\tilde{\Gamma}$. 

Remove a regular neighborhood of $\Gamma$ to define the manifold with boundary $M = M_E - N(\Gamma)$.\footnote{3D gravity lives on $M$, see \cite{gravitypaper}.} The triangulation $T$ provides a marked Heegaard splitting $(C,\gamma,H, \Sigma)$ of $M$, split along $\Sigma = \p  N(T)$. The compression body $C$ in this splitting is defined by starting with the Riemann surface $\p N(T)$ and filling in the cycles in $\tilde{\Gamma}$, and the marking $\gamma$ consists of the meridian cycles of the edges in $\Gamma$. The handlebody is $H = M_E - N(T)$. 

Therefore, given a VTQFT graph $(S^3, \Gamma(\vecP))$, we have obtained a marked Heegaard diagram $(\Sigma, \gamma, \tilde{\gamma}, \rho)$, with edges in $\Gamma$ corresponding to the marking $\gamma$. Embed this diagram in $S^3$, and push the curves $\gamma$ and $\tilde{\gamma}$ slightly inward, so that they are enclosed by the curves $\rho$. Assign framings parallel to $\Sigma$. Finally, label the curves $\gamma$ by $\vecP$, and all other curves by $\Omega$. This defines the chain-mail link $\Gamma_{CH}(\vecP)$, and thereby gives a second definition of the chain-mail invariant, $\langle \Gamma_{CH}(\vecP)\rangle$.

The Heegaard splitting is of course not unique. Any marked Heegaard diagram for $M$ defines a chain-mail link in the same way, and the link depends on the choice of splitting. However, it was proved in \cite{roberts1995skein,Barrett:2004im} that the amplitude $\langle \Gamma_{CH}(\vecP)\rangle$ is independent of the Heegaard splitting. The proof carries over verbatim to the Virasoro case.

The diagrammatic rules \eqref{cmrules}  were designed to produce a marked Heegaard diagram for $M$. It follows that the chain-mail invariant defined diagrammatically in section \ref{ss:chainmailGraph} is the same as the chain-mail invariant defined from a Heegaard splitting. For example, let us redraw the link $\Gamma_{CH}$ in \eqref{gammaCHhandcuff}, first with color-coding to help keep track of the cycles, and then on a genus-4 surface:
\begin{align}\label{genus4heegaard}
\cp{\includegraphics[width=2.5in]{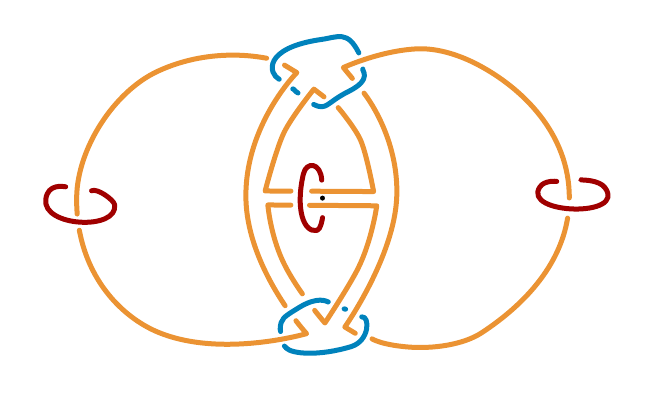}}
\longrightarrow
\cp{\includegraphics[width=3in]{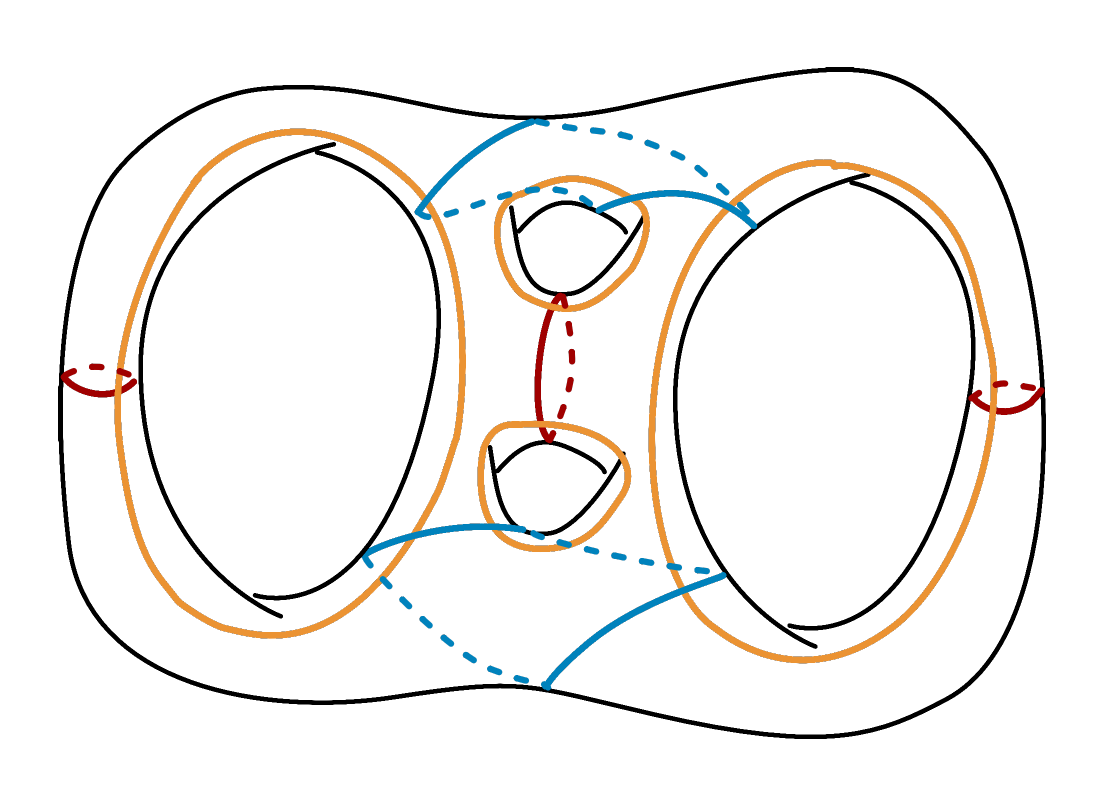}}
\end{align}
This is the marked Heegaard diagram for a genus-4 splitting of $M$. The red loops are the marking $\gamma(\vecP)$, the blue loops are the cycles $\tilde{\gamma}$ contractible in the compression body, and the orange loops are the cycles $\rho$ contractible in the handlebody. This generalizes immediately to any chain-mail diagram produced by the rules in section \ref{ss:chainmailGraph}.

\subsection{Chain mail and the CTV partition function}

Given a graph $(S^3, \Gamma(\vecP))$ and a triangulation, the marked Heegaard diagram inherited from the triangulation can also be described as follows. The handlebody in the Heegaard splitting is the thickened dual complex, and we will draw the marked Heegaard diagram on the surface of this handlebody. The piece of the handlebody coming from a single tetrahedron is
\begin{align}\label{dualhb}
\cp{\includegraphics[width=2in]{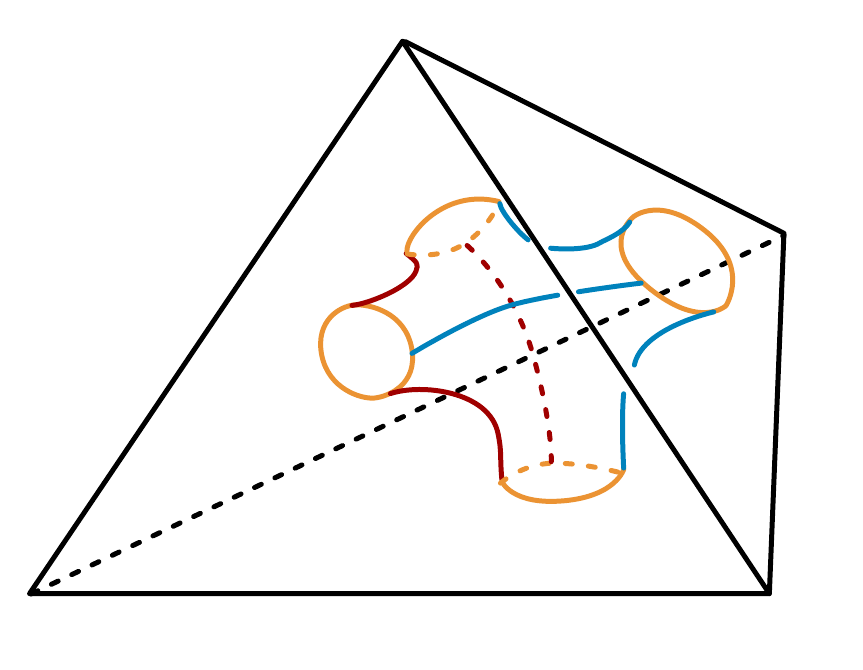}}
\end{align}
This example has three internal edges (blue) and three external edges (red). The orange cycles are contractible in the handlebody, so these are the cycles in $\rho$. The red and blue cycles, which correspond to edges of the tetrahedron, belong to $\gamma$ if the edge is external and $\tilde{\gamma}$ if the edge is internal. The marked Heegaard diagram is obtained by repeating this for each tetrahedron and gluing them together. Therefore, the chain-mail link $\Gamma_{CH}(\vecP)$ locally takes the form
\begin{align}
\cp{\includegraphics[width=1.5in]{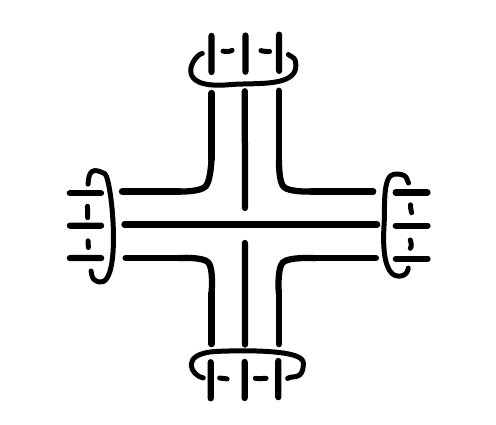}}
\end{align}
around each tetrahedron. To evaluate the chain-mail invariant $\Zvir(S^3, \Gamma_{CH}(\vecP))$, we use the vertex identity \eqref{vertexjoin} three times to pinch off three of the loops, giving
\begin{align}
\cp{\includegraphics[width=1.3in]{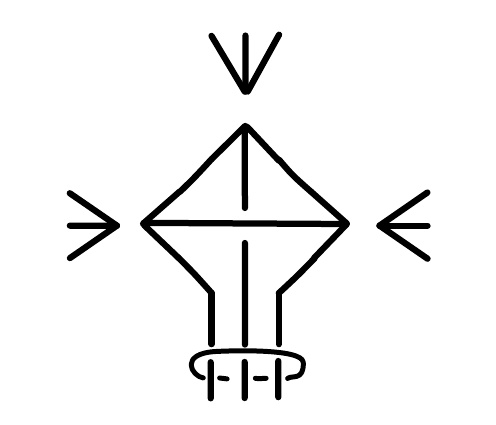}}
\end{align}
The remaining triangle produces a $6j$-symbol using \eqref{triangleremoval}. 
This demonstrates that each tetrahedron in the triangulation contributes as in \eqref{tetW}, and we recall that internal edges are assigned the weight $\Omega$ and thus integrated with the Cardy measure. Therefore, the chain-mail invariant is equal to the CTV partition function \eqref{defCTV}:
\begin{align}
\Zvir(S^3, \Gamma_{CH}(\vecP)) = \Ztv(S^3, \Gamma(\vecP)) \ . 
\end{align}
To understand the restriction to large triangulations, suppose we ignore it temporarily, and try to apply this procedure to a triangulation that includes four or more tetrahedra meeting at an internal vertex, as in \eqref{badvertex}.
The statement that the centroid vertex is internal means that the weights on all four incident edges are integrated. It is straightforward to check that the resulting chain-mail link for this part of the triangulation has a contractible $\Omega$-loop, coming from one of the internal edges that separates from the rest of the graph. This diverges in Virasoro TQFT, so if there are vertices of this type then the CTV partition function is ill defined. This is in contrast to a discrete spin network, where the integrated bubble would produce a harmless factor of the total quantum dimension.

\subsubsection*{Revisiting the modular $S$-matrix example}
Let us now describe the example of the modular $S$-matrix in this language. In section \ref{sss:ctvSmatrix} we calculated the CTV partition from a triangulation with a single tetrahedron, and compared it to $|\Zvir|^2$. Then, around \eqref{genus4heegaard} we derived the chain-mail link and demonstrated that it gives the marked Heegaard diagram for a genus-4 splitting of $(S^3, \Gamma(\vecP))$.

The genus-4 splitting in \eqref{genus4heegaard} is not the splitting associated to the triangulation in section \ref{sss:ctvSmatrix}, which we will now describe. This triangulation has a single tetrahedron, so we have a single contribution like \eqref{dualhb} with all edges external. Gluing the ends as dictated by the triangulation, the handlebody is
\begin{align}\label{genus2heegaard}
\cp{\includegraphics[width=1.4in]{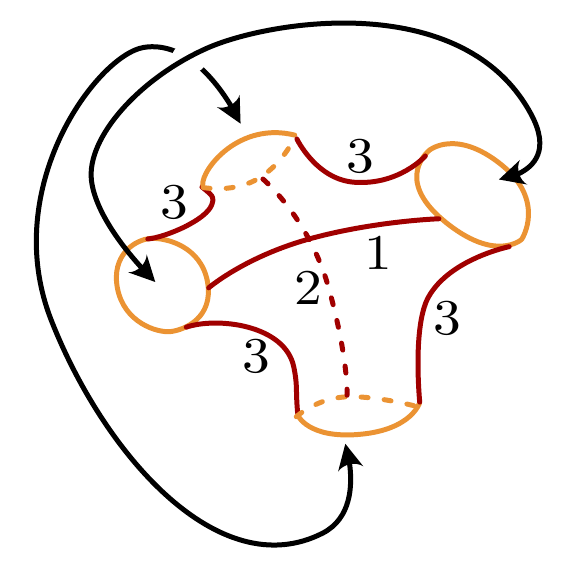}}
\qquad = \qquad
\cp{\includegraphics[width=2.2in]{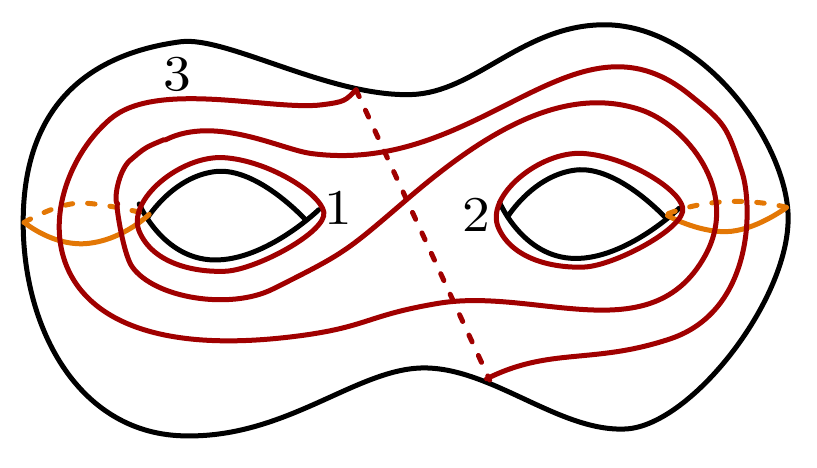}}
\end{align}
This is the marked Heegaard diagram for a genus-2 splitting of $(S^3,\Gamma(\vecP))$. The chain-mail amplitude is the same for \eqref{genus4heegaard} and \eqref{genus2heegaard}.

\section{Shadow formalism}
The derivation of the main identity \eqref{introFourierVV} requires one more ingredient, a technique to evaluate TQFT graphs known as the shadow formalism \cite{MR1026957,MR1280463,Turaev:1992hq,Turaev:1994xb}. This is well known in the TQFT/spin network literature, but it has not been discussed in Virasoro TQFT, and requires a few small changes as compared to these references to avoid divergences. We will therefore review/extend the shadow technique applied to Virasoro TQFT, before returning to the discussion of chain mail in the next section.

The shadow technique can be understood geometrically in terms of dual spines, but we do not need that perspective here. We will present it as a set of diagrammatic rules to evaluate $\Zvir(M_E, \Gamma(\vecP))$. As explained in section \ref{s:vtqft}, we already have a set of rules to evaluate graphs --- so why do we need another? One answer is that the shadow evaluation is what naturally appears in the derivation of the main identity below. But a deeper answer is that the shadow formalism is a way to express the VTQFT amplitude in a way that has a clear and direct relationship to geometry. The rules in section \ref{s:vtqft} reduce the amplitude to an integral of a product of $6j$-symbols and phases --- it is the phases that are difficult to interpret geometrically. The shadow evaluation gives the amplitude as an integral of tetrahedral building blocks without extra phases.

We will state the general rules and simultaneously work out the example of the knotted handcuff graph \eqref{gammahandcuff}. For the derivation we refer to \cite[Chapter 11]{MR1280463}. The rules are applied to the graph diagram projected onto the plane in the usual way with under- and over-crossings. The first step is to shade the planar regions of the graph in gray and white, with white on the outermost region, and isotope the graph so that all crossings are shaded gray in the north/south regions. That is, all crossings must be shaded as 
\begin{align}
\cp{\begin{overpic}[percent,width=1.5in]{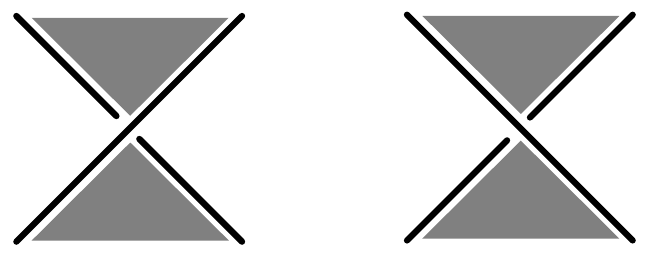}
\put (45,20) {or}
\end{overpic}
} \ . 
\end{align}
For example, the graph in \eqref{gammahandcuff} can be shaded as
\begin{align}\label{shadedhandcuff}
\cp{\includegraphics[width=0.6in]{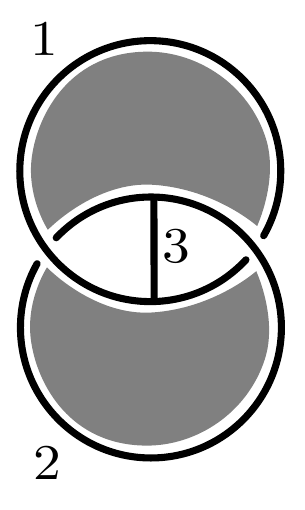}} \ . 
\end{align}
Once the graph is drawn this way, we can ignore the shadings, so we will not continue to draw them. Next, we $(i)$ project the graph onto the plane; $(ii)$ label each planar region of the projected graph by a conformal weight $P_i$, with $\id$ on the outermost region; and $(iii)$ depict the edges as dashed lines and the crossings as
\begin{align}
\cp{\includegraphics[width=1.5in]{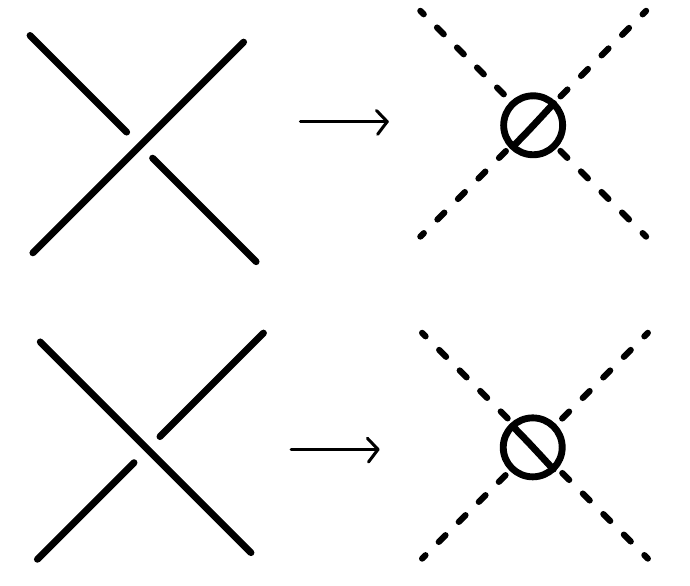}}
\end{align}
This produces the shadow diagram, in which each edge and each planar region is labeled by a conformal weight. For our example \eqref{shadedhandcuff}, the shadow diagram is
\begin{align}\label{hcshadow}
\cp{\begin{overpic}[percent,width=1.9in]{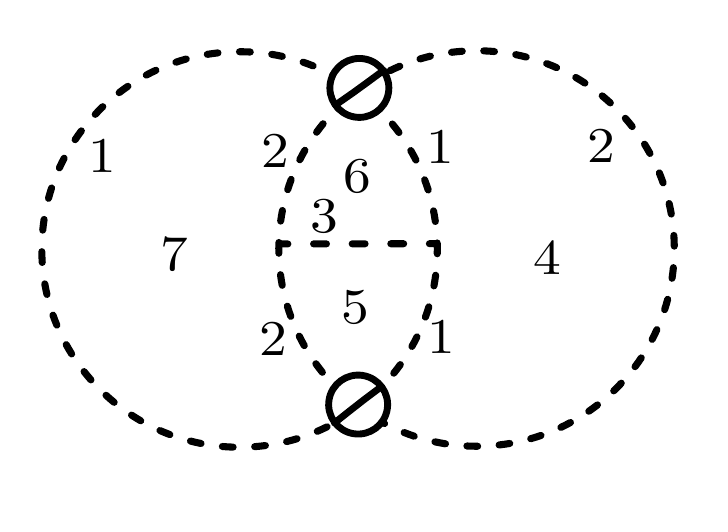}
\put (90,10) {$\id$}
\end{overpic}}
\end{align}
In the `shadow world', the graph $\Gamma$ has become purely 2-dimensional. 

The amplitude $\Zvir(S^3, \Gamma(\vecP))$ can be read off immediately from the shadow diagram. Each crossing or vertex produces a $6j$ factor as follows:
\begin{align}
\cp{\includegraphics[width=0.8in]{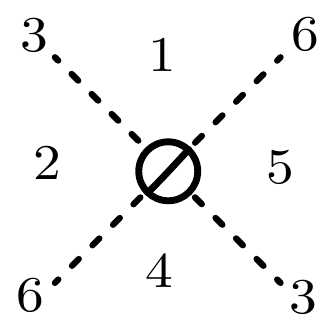}}
\qquad &\longrightarrow \qquad
\vbracket{\footnotesize
\begin{tikzpicture}[scale=1.3]
\coordinate (v1) at (0,1) ;
\coordinate (v2) at (1,1); 
\coordinate (v3) at (0,0); 
\coordinate (v4) at (1,0); 
\draw[very thick] (v1) -- (v2) node[above,midway] {$1$};
\draw[very thick]  (v1) -- (v3) node[left,midway] {$2$};
\draw[very thick]  (v3) -- (v4) node[below,midway] {$4$};
\draw[very thick]  (v4) -- (v2) node[right,midway] {$5$};
\draw[very thick]  (v3) -- (v2) node[pos=0.2,above] {$6$};
\draw[line width=4pt,white,shorten <= 3pt,shorten >= 3pt] (v1) -- (v4) ;
\draw[very thick]  (v1) -- (v4) node[pos=0.8,above] {$3$};
\end{tikzpicture}
}
 \\
\cp{\includegraphics[width=0.8in]{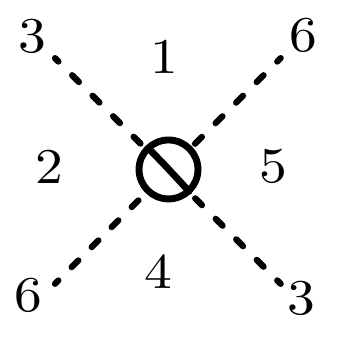}}
\qquad &\longrightarrow \qquad 
\vbracket{\footnotesize
\begin{tikzpicture}[scale=1.3]
\coordinate (v1) at (0,1) ;
\coordinate (v2) at (1,1); 
\coordinate (v3) at (0,0); 
\coordinate (v4) at (1,0); 
\draw[very thick] (v1) -- (v2) node[above,midway] {$1$};
\draw[very thick]  (v1) -- (v3) node[left,midway] {$2$};
\draw[very thick]  (v3) -- (v4) node[below,midway] {$4$};
\draw[very thick]  (v4) -- (v2) node[right,midway] {$5$};
\draw[very thick]  (v1) -- (v4) node[pos=0.8,above] {$3$};
\draw[line width=4pt,white,shorten <= 3pt,shorten >= 3pt] (v3) -- (v2) ;
\draw[very thick]  (v3) -- (v2) node[pos=0.2,above] {$6$};
\end{tikzpicture}
}
\\
\cp{\includegraphics[width=0.7in]{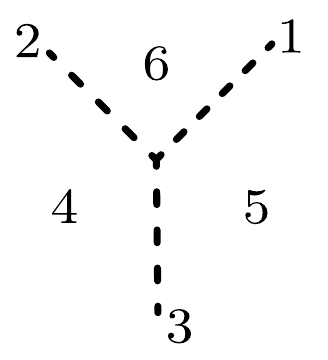}}
\qquad &\longrightarrow \qquad 
 \vbracket{\footnotesize
\begin{tikzpicture}[scale=0.8]
\centerarc[very thick](0,0)(-30:90:1);
\centerarc[very thick](0,0)(90:210:1);
\centerarc[very thick](0,0)(210:335:1);
\draw[very thick] (0,0) -- (0,1);
\draw[very thick] (0,0) -- ({cos(30)},{-sin(30)});
\draw[very thick] (0,0) -- ({-cos(30)},{-sin(30)});
\node at (.9,.9) {$1$};
\node at (-.9,.9) {$2$};
\node at (-0.2,.55) {$3$};
\node at (-.55,.0) {$4$};
\node at (0.55,0) {$5$};
\node at (0,-1.25) {$6$};
\end{tikzpicture}}
\end{align}
and internal weights are integrated with the Cardy measure. For example, the shadow diagram \eqref{hcshadow} evaluates to
\begin{align}\label{hcshadoweval}
\vbracket{\includegraphics[width=1.1in]{figures/handcuff.pdf}}
&= \int_{4567}
 \vbracket{\footnotesize
\begin{tikzpicture}[scale=0.8]
\centerarc[very thick](0,0)(-30:90:1);
\centerarc[very thick](0,0)(90:210:1);
\centerarc[very thick](0,0)(210:335:1);
\draw[very thick] (0,0) -- (0,1);
\draw[very thick] (0,0) -- ({cos(30)},{-sin(30)});
\draw[very thick] (0,0) -- ({-cos(30)},{-sin(30)});
\node at (.9,.9) {$1$};
\node at (-.9,.9) {$1$};
\node at (-0.2,.55) {$3$};
\node at (-.55,.0) {$6$};
\node at (0.55,0) {$5$};
\node at (0,-1.25) {$4$};
\end{tikzpicture}}
 \vbracket{\footnotesize
\begin{tikzpicture}[scale=0.8]
\centerarc[very thick](0,0)(-30:90:1);
\centerarc[very thick](0,0)(90:210:1);
\centerarc[very thick](0,0)(210:335:1);
\draw[very thick] (0,0) -- (0,1);
\draw[very thick] (0,0) -- ({cos(30)},{-sin(30)});
\draw[very thick] (0,0) -- ({-cos(30)},{-sin(30)});
\node at (.9,.9) {$2$};
\node at (-.9,.9) {$2$};
\node at (-0.2,.55) {$3$};
\node at (-.55,.0) {$5$};
\node at (0.55,0) {$6$};
\node at (0,-1.25) {$7$};
\end{tikzpicture}}
\vbracket{\footnotesize
\begin{tikzpicture}[scale=1.3]
\coordinate (v1) at (0,1) ;
\coordinate (v2) at (1,1); 
\coordinate (v3) at (0,0); 
\coordinate (v4) at (1,0); 
\draw[very thick] (v1) -- (v2) node[above,midway] {$7$};
\draw[very thick]  (v1) -- (v3) node[left,midway] {$6$};
\draw[very thick]  (v3) -- (v4) node[below,midway] {$4$};
\draw[very thick]  (v4) -- (v2) node[right,midway] {$\id$};
\draw[very thick]  (v1) -- (v4) node[pos=0.8,above] {$2$};
\draw[line width=4pt,white,shorten <= 3pt,shorten >= 3pt] (v3) -- (v2) ;
\draw[very thick]  (v3) -- (v2) node[pos=0.2,above] {$1$};
\end{tikzpicture}
}
\vbracket{\footnotesize
\begin{tikzpicture}[scale=1.3]
\coordinate (v1) at (0,1) ;
\coordinate (v2) at (1,1); 
\coordinate (v3) at (0,0); 
\coordinate (v4) at (1,0); 
\draw[very thick] (v1) -- (v2) node[above,midway] {$7$};
\draw[very thick]  (v1) -- (v3) node[left,midway] {$\id$};
\draw[very thick]  (v3) -- (v4) node[below,midway] {$4$};
\draw[very thick]  (v4) -- (v2) node[right,midway] {$5$};
\draw[very thick]  (v1) -- (v4) node[pos=0.8,above] {$1$};
\draw[line width=4pt,white,shorten <= 3pt,shorten >= 3pt] (v3) -- (v2) ;
\draw[very thick]  (v3) -- (v2) node[pos=0.2,above] {$2$};
\end{tikzpicture}
}\notag
\\
&= \int_{56} \begin{Bmatrix}
P_3 & P_5& P_6 \\
P_2 & P_1 & P_1 \end{Bmatrix}
\begin{Bmatrix}
P_3 & P_5 & P_6\\
P_1 & P_2 & P_2
\end{Bmatrix}
\braidB_{6}^{12}\braidB_{5}^{12}  
\end{align}
It is straightforward to check that the last expression equals the expected answer, $\hatS_{P_1P_2}[P_3]$, using the usual VTQFT methods.\footnote{First draw the product of two 6j-symbols in the last line of \eqref{hcshadoweval} as a triangle inside a tetrahedron, similar to the figure in \eqref{trisq} but with different labels. Then absorb the braiding factors in \eqref{hcshadoweval} into this diagram by braiding two of the vertices, and do the integrals over $5,6$ using \eqref{identityfusion}. The result is the original graph.}

\section{Relating Chain Mail to $|\Zvir|^2$}\label{s:mainidentity}
We will now show that the chain-mail invariant is related to $|\Zvir|^2$ by an $S$-transform,
\begin{align}
\Zvir(M_E, \Gamma_{CH}(\vecP)) = 
 \int_{\mathbb{R}_+^n}d\vecP (\Pi_{i=1}^n S_{P_i' P_i})
\left|
\Zvir(M_E, \Gamma(\vecP))
\right|^2
\end{align}
The derivation follows \cite{roberts1995skein,Barrett:2004im} very closely, so this is mostly review, aside from the minor changes to the definition of the chain-mail link --- the  nontrivial step to extend these results to Virasoro was showing that the chain-mail formalism can be adapted to this context, which we have already done above.

\subsection{Graphs in $S^3$}

We use the diagrammatic definition of the chain-mail invariant from section \ref{ss:chainmailGraph}. Given a graph $\Gamma(\vecP)$, the chain-mail link  $\Gamma_{CH}(\vecP)$ was defined diagramatically. Write the Fourier transform of the chain-mail invariant as
\begin{align}
I(\Gamma(\vecP')) = \int_{\mathbb{R}_+^n}(\Pi_{i=1}^ndP_i S_{P_i'P_i}) \Zvir(S^3, \Gamma_{CH}(\vecP))
\end{align}
Recall that the external edges in the chain-mail link, labeled by $\vecP$, appear as
\begin{align}
\cp{\footnotesize
\begin{overpic}[percent,width=1in]{figures/sloop.pdf}
\put (0,30) {\footnotesize $\Omega$}
\put (60,40) {\footnotesize $i$}
\end{overpic}
}
\qquad \mbox{or} \qquad
\cp{\footnotesize
\begin{overpic}[percent,width=1in]{figures/doubleloop.pdf}
\put (10,4) {$\Omega$}
\put (10,35) {$\Omega$}
\put (60,45) {$i$}
\end{overpic}
}
\end{align}
Therefore the Fourier transform $I(\Gamma(\vecP'))$ can be calculated by replacing 
\begin{align}\label{ftrepsCH}
\cp{\footnotesize
\begin{overpic}[percent,width=1in]{figures/sloop.pdf}
\put (0,30) {\footnotesize $\Omega$}
\put (60,40) {\footnotesize $i$}
\end{overpic}
} &\qquad \longrightarrow \qquad
\cp{\graphEdge{$i'$}} \\
\cp{\footnotesize
\begin{overpic}[percent,width=1in]{figures/doubleloop.pdf}
\put (60,45) {$i$}
\end{overpic}
}
&\qquad \longrightarrow
\qquad
\cp{\footnotesize
\begin{overpic}[percent,width=1.2in]{figures/doublelooplink.pdf}
\put (70, 55) {$i'$}
\put (33,46) {$\Omega$}
\end{overpic}
}
\end{align}
everywhere in $\Gamma_{CH}(\vecP)$. By \eqref{doublelooplink}, the second line is equivalent to
\begin{align}
\cp{\footnotesize
\begin{overpic}[percent,width=1in]{figures/doubleloop.pdf}
\put (60,45) {$i$}
\end{overpic}
}
\qquad \longrightarrow
\qquad
\cp{\graphSbig{}{}{}{}{i'}}
\end{align}
After making this replacement in $\Gamma_{CH}(\vecP)$, we evaluate the VTQFT amplitude using the identities
\begin{align}\label{replace3}
\vbracket{\footnotesize\begin{overpic}[percent,width=1in]{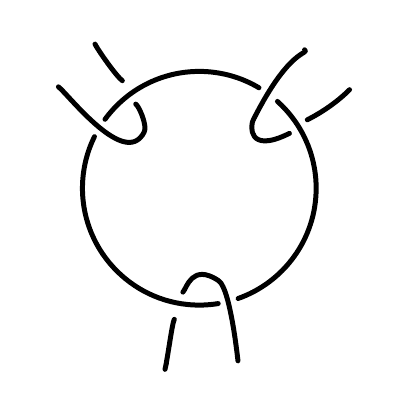}
\put (5,80) {1}
\put (75,85) {2}
\put (55,5) {3}
\put (72,30) {$\Omega$}
\end{overpic}}
\quad=\quad
\vbracket{\footnotesize\begin{overpic}[percent,width=1in]{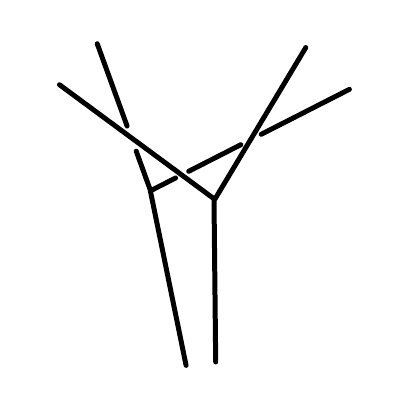}
\put (5,80) {1}
\put (20,92) {1}
\put (72,92) {2}
\put (84,80) {2}
\put (37,6) {3}
\put (53,6) {3}
\end{overpic}} e^{i\pi(h_1+h_2+h_3)}
\end{align}
and
\begin{align}\label{replace4}
\vbracket{\footnotesize\begin{overpic}[percent,width=1in]{figures/chaincross2.pdf}
\put (8,68) {1}
\put (22,80) {2}
\put (67,80) {2}
\put (85,68) {3}
\put (88,11) {3}
\put (75,-3) {4}
\put (17,-3) {4}
\put (6,9) {1}
\put (47,10) {$\Omega$}
\end{overpic}}
\quad=\quad
\int_{56}
\vbracket{\footnotesize\begin{overpic}[percent,width=1in]{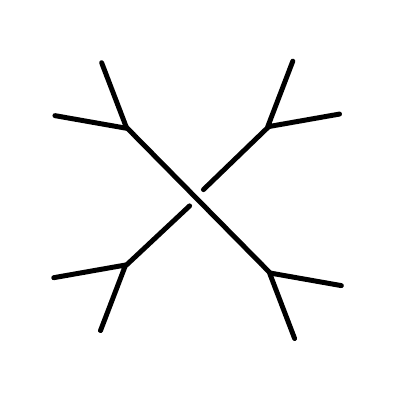}
\put (7,68) {1}
\put (22,87) {2}
\put (67,87) {2}
\put (85,68) {3}
\put (85,20) {3}
\put (68,8) {4}
\put (17,10) {4}
\put (6,30) {1}
\put (38,63) {5}
\put (31,45) {6}
\end{overpic}
}
\vbracket{\footnotesize\begin{overpic}[percent,width=0.8in]{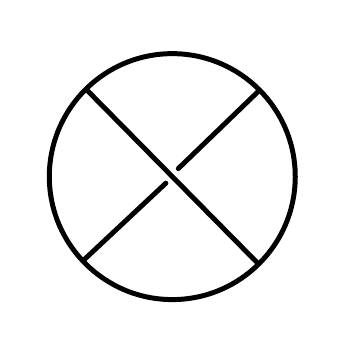}
\put (6,44) {1}
\put (86,44) {3}
\put (46,89) {2}
\put (46,3) {4}
\put (35,65) {5}
\put (57,65) {6}
\end{overpic}
}
\end{align}
The first is a braided version of \eqref{vertexjoin}, and the second is a fusion transformation of \eqref{quadvertex} followed by braiding. Finally, remove bubbles using \eqref{bubbleremoval}. The final result is the original graph $\langle \Gamma(\vecP) \rangle$ times a product of $6j$-symbols coming from \eqref{replace3} and \eqref{replace4}. This product of $6j$-symbols is manifestly equal to $\langle \bar{\Gamma}(P)\rangle$, the amplitude for the graph where over-crossings are replaced by under-crossings and vice-versa, when this amplitude is evaluated by the shadow method. Hence
\begin{align}\label{igammasq}
I(\Gamma(\vecP')) &= | \langle \Gamma(\vecP')\rangle|^2
\end{align}
which is what we wanted to show.

As an example, consider the knotted handcuff graph \eqref{gammahandcuffB}.  The chain-mail link is \eqref{gammaCHhandcuff}. Its Fourier transform is calculated by applying \eqref{ftrepsCH}, which gives
\begin{align}
I(\Gamma(\vecP')) \quad=\quad
\int_{45}
\vbracket{\footnotesize
\begin{overpic}[width=2in,percent]{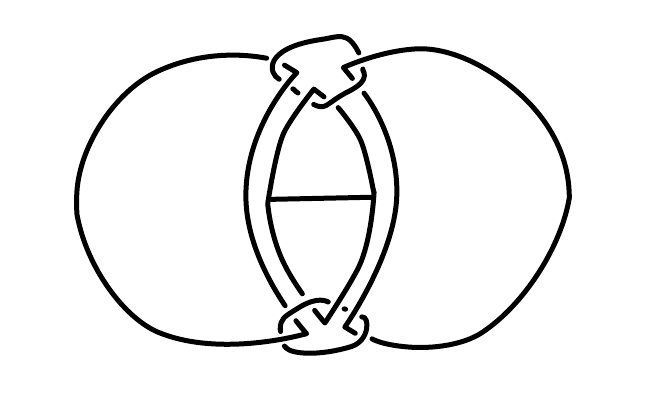}
\put (6,29) {$1'$}
\put (90,29) {$2'$}
\put (49,32) {$3'$}
\put (45,38) {$4$}
\put (45,20) {$5$}
\put (45,57) {$\Omega$}
\put (48,1) {$\Omega$}
\end{overpic}
}
\end{align}
Using \eqref{replace3} then \eqref{triangleremoval},
\begin{align}
I(\Gamma(\vecP')) &= \int_{45}
\vbracket{\footnotesize\begin{overpic}[width=2in,percent]{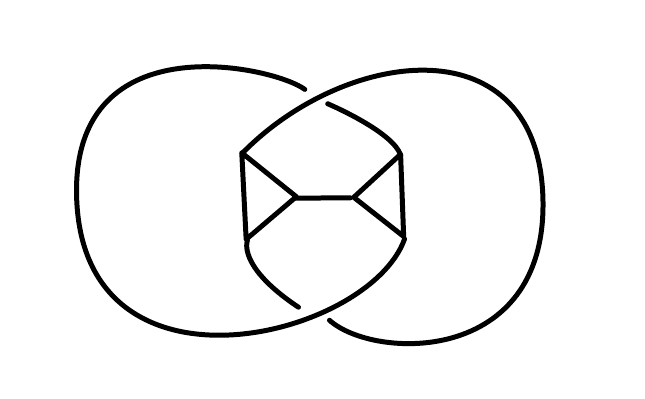}
\put (6,29) {$1'$}
\put (86,29) {$2'$}
\put (49,33) {$3'$}
\put (42,35) {$4$}
\put (55,35) {$4$}
\put (42,23) {$5$}
\put (55,23) {$5$}
\put (32,29) {$1'$}
\put (65,29) {$2'$}
\end{overpic}}
e^{i\pi (2h_1'+2h_2' - h_4 -h_5)}
\\
&= \quad
\vbracket{\footnotesize\begin{overpic}[percent,width=1.3in]{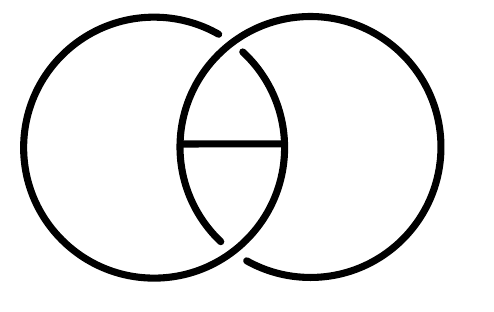}
\put (10,30) {$1'$}
\put (80,30) {$2'$}
\put (45,36) {$3'$}
\end{overpic}
}
\int_{45} \begin{Bmatrix} P_2' & P_2' & P_3' \\
P_4 & P_5 & P_1'
\end{Bmatrix}
\begin{Bmatrix} P_1' &P_1' & P_3' \\
P_4 & P_5 & P_2'
\end{Bmatrix} e^{i\pi (2h_1'+2h_2' - h_4 -h_5)}
\end{align}
The integral is equal to $\hatS_{P_1'P_2'}^*[P_3']$, as discussed around \eqref{hcshadoweval}. This confirms \eqref{igammasq} in this example.

\subsection{More general embedding manifolds}\label{ss:generalME}
Thus far, our discussion of the chain-mail invariant has been restricted to graphs in $ S^3$. The extension to graphs in a more embedding manifold $M_E$ is straightforward. The Lickorish-Wallace theorem states that any closed, orientable, connected 3-manifold can be obtained by Dehn surgery on a knot or link in $S^3$ \cite{MR125588,MR151948}.\footnote{This result has been invoked in AdS$_3$ gravity and VTQFT --- for similar reasons, namely to change the topology of the embedding manifold --- in the recent work \cite{Jafferis:2025vyp}.} Therefore, given a surgery presentation of $M_E$, the VTQFT amplitude, CTV partition function, and chain-mail invariant on $(M_E, \Gamma(\vecP))$ can all be recast as amplitudes on $(S^3, \hat{\Gamma}(\vecP))$ where $\hat{\Gamma}$ is an extension of $\Gamma$ by the surgery link. This extends our main results --- the chain-mail formula for the CTV partition function \eqref{ctvcm} and the main identity \eqref{introFourierVV} relating CTV to VTQFT --- to graphs embedded in a closed orientable 3-manifold $M_E$, whenever the VTQFT amplitude and CTV partition function are well defined. We will not attempt to characterize the admissible $(M_E, \Gamma)$, which is an interesting question for the future, but setting aside this caveat, the derivations of the main results in $M_E$ are identical to \cite{roberts1995skein,Barrett:2004im}.

\ \bigskip

\noindent\textbf{Acknowledgments}

\noindent  It is a pleasure to thank Alex Belin, Jeevan Chandra, Wan Zhen Chua, Scott Collier, Gabriele Di Ubaldo, Luca Iliesiu, Yikun Jiang, Adam Levine, Nate MacFadden, Juan Maldacena, Alex Maloney, Greg Mathys, Eric Perlmutter, Julian Sonner, Douglas Stanford, Diandian Wang, and Cynthia Yan for discussions on related topics. This work is supported by NSF grant PHY-2309456.

\appendix

\section{Virasoro crossing kernels}\label{app:kernels}
In this appendix we review the transformations of Virasoro conformal  blocks \cite{Ponsot:2000mt,Teschner:2012em,Teschner:2013tqy, Eberhardt:2023mrq}. We adopt the conventions and notation for conformal blocks and crossing kernels from the comprehensive review \cite{Eberhardt:2023mrq} as well as \cite{Collier:2023fwi,Collier:2024mgv} with one exception: The letter $h$ is used for a chiral conformal weight, reserving $\Delta$ for the total scaling weight, $\Delta = h + \bh$.  The weights and central charge are parameterized as
\begin{align}
h = \frac{Q^2}{4}  + P^2 , \quad c = 1+6Q^2 , \quad Q = b + b^{-1}
\end{align}
with $b \to 0$ in the semiclassical limit. It is also common to use
\begin{align}
h = \alpha (Q - \alpha) \ , \quad \alpha = \frac{Q}{2} + i P   = \frac{\eta}{b} \ .
\end{align}
We assume all non-vacuum weights satisfy $h \geq \frac{c-1}{24}$,  $P \in \mathbb{R_+}$ unless otherwise noted. 
We will sometimes use the shorthand $P_1 \to 1$, $P_2 \to 2$, etc., as in $\fusionF{56}{2&1\\3&4} \equiv \fusionF{P_5P_6}{P_2 & P_1\\ P_3 & P_4}$ and $C_0(123) \equiv C_0(P_1,P_2,P_3)$.

\subsection{Fusion kernel and $C_0$}
The fusion transformation for Virasoro conformal blocks is
\begin{align}\label{fusionaa}
\graphS{3}{4}{2}{1}{P}
= \int_0^\infty dR\, \fusionF{PR}{3& 2\\4& 1} 
\graphT{3}{4}{2}{1}{R} \ , 
\qquad
\graphT{3}{4}{2}{1}{R} = \int_0^\infty dP\, 
\fusionF{RP}{2& 1\\3& 4}\graphS{3}{4}{2}{1}{P} \ . 
\end{align}
This can also be applied to internal lines of conformal blocks for higher-point functions on a genus-$g$ Riemann surface. The fusion kernel squares to one, in the sense
\begin{align}\label{fusionsq}
\int_0^\infty dR\, 
\fusionF{PR}{3& 2\\ 4& 1}
\fusionF{RP'}{4& 3\\ 1& 2}
= \delta(P-P')
\end{align}
for $P,P' > 0$. 
If the external operators are pairwise identical then the identity block appears in the OPE. The fusion kernel for the identity block plays a special role as it defines the function $C_0$ \cite{Collier:2019weq},
\begin{align}\label{c0fromF}
C_0(123) \equiv C_0(P_1,P_2,P_3) &= \frac{1}{\rho_0(P_3)} \fusionF{\id 3}{2 & 1\\ 2& 1} \ ,
\end{align}
with
\begin{align}\label{fusionvac}
\graphS{2}{2}{1}{1}{\id}
= \int_0^\infty dP\, \rho_0(P) C_0(12P)
\graphT{2}{2}{1}{1}{P}  \ . 
\end{align}
$C_0$ is real, and symmetric in its arguments. Under analytic continuation $P_3 \to \id$ it satisfies
\begin{align}
C_0(P_1, P_2, \id) = \frac{1}{\rho_0(P_1)} \delta(P_1 - P_2) 
\end{align}
The Virasoro $6j$-symbol in Racah-Wigner normalization is 
\begin{align}
\begin{Bmatrix}
P_1 & P_2 & P_3 \\ P_4 & P_5 & P_6
\end{Bmatrix}
&=
\frac{1}{\rho_0(P_6)} \sqrt{\frac{C_0(123)C_0(345)}{C_0(156)C_0(246)}} \fusionF{36}{4& 2\\ 5& 1}
\end{align}
This has tetrahedral symmetry, generated by column permutations and
$
\begin{Bsmallmatrix}
P_1 & P_2 & P_3 \\ P_4 & P_5 & P_6
\end{Bsmallmatrix} = 
\begin{Bsmallmatrix}
P_4 & P_5 & P_3 \\ P_1 & P_2 & P_6
\end{Bsmallmatrix}
$.
The identity fusion kernel is
\begin{align}
\begin{Bmatrix}P_1&P_2&\id\\P_3&P_4&P_5\end{Bmatrix}
&= \frac{ \delta(P_1-P_2)\delta(P_3-P_4)}{ \sqrt{ \rho_0(P_1)\rho_0(P_3) }}
\end{align}
The orthogonality relation \eqref{fusionsq} is equivalent to
\begin{align}\label{sixjortho}
\int_0^\infty dR\, \rho_0(R) \begin{Bmatrix} P_1 & P_2 & P \\ P_3 & P_4 & R \end{Bmatrix}
\begin{Bmatrix} P_2 & P_3 & R \\ P_4 & P_1 & P' \end{Bmatrix}
&= \frac{1}{\rho_0(P)} \delta(P-P')
\end{align}
The hexagon identity in Racah-Wigner normalization is 
\begin{align}
\begin{Bmatrix}
P_1 & P_2 & P_3 \\
P_4 & P_5 & P_6
\end{Bmatrix}
&=
\int_0^\infty dP_0 \, \rho_0(P_0)
e^{i\pi( h_1+h_2+h_4+h_5 - h_3-h_6-h_0)}
\begin{Bmatrix}
P_1& P_4 & P_0\\
P_2 & P_5 & P_6 
\end{Bmatrix}
\begin{Bmatrix}
P_2 & P_1 & P_3 \\
P_4 & P_5 & P_0
\end{Bmatrix}
\end{align}
The pentagon identity is
\begin{align}
\int_0^\infty dP\, \rho_0(P) 
\begin{Bmatrix} a_2 & a_1 & c_3 \\ b_1 & b_2 & P \end{Bmatrix}
\begin{Bmatrix} a_3 & a_1 & c_2 \\ b_1 & b_3 & P \end{Bmatrix}
&\begin{Bmatrix} a_3 & a_2 & c_1 \\ b_2 & b_3 & P \end{Bmatrix}\\
&\qquad
=
\begin{Bmatrix} c_1 & c_2 & c_3 \\ b_1 & b_2 & b_3 \end{Bmatrix}
\begin{Bmatrix} c_1 & c_2 & c_3 \\ a_1 & a_2 & a_3 \end{Bmatrix}\notag
\end{align}
This corresponds to the $3-2$ Pachner move, which decomposes two tetrahedra, glued along one face, into three tetrahedra glued around an internal edge. 

For the integral representation of $\mathbb{F}$ or $\begin{Bsmallmatrix} P_1 & P_2 & P_3 \\ P_4 & P_5 & P_6 \end{Bsmallmatrix}$ see \cite{Ponsot:2000mt, Teschner:2012em}, or the review \cite[eqn (3.50)]{Eberhardt:2023mrq}.



\subsection{Modular S-matrix}
The one-point conformal block on the torus with external weight $R$ and internal weight $P$ is written as 
$\torusBlock{P}{R}$
and its modular $S$-transformation under $\tau \to -1/\tau$ is written $\torusBlockS{P}{R}$. This transformation defines the Virasoro modular S-matrix $\modularS_{P_1P_2}[P_3]$,
\begin{align}
\torusBlock{P}{1}
&= \int_0^\infty dR\ \modularS_{PR}[P_1] \torusBlockS{R}{1} \quad \ . 
\end{align}
The phase is fixed by
\begin{align}\label{sphase}
\modularS_{P_1P_2}[P_3] &= e^{i\pi h_3/2} \times \mbox{(real)} \ ,
\end{align}
assuming $P_i \in \mathbb{R}_+$. 
The $PSL(2,Z)$ relation $S^2=1$ implies
\begin{align}
\int_0^\infty dQ \, \modularS_{PQ}[R] \modularS_{QP'}[R]^* = \delta(P-P') \ .
\end{align}
and $(ST)^3=1$ results in 
\begin{align}\label{stcubed}
\int_0^\infty dQ \, \modularS_{PQ}[P_3] \modularS_{QR}[P_3] e^{-2\pi i(h_P+h_Q+h_R-c/8)} 
&= \modularS_{PR}[P_3] \ . 
\end{align}
The modular transformation of the vacuum character defines the (chiral) Cardy density of states $\rho_0$,
\begin{align}
 \rho_0(P) &=\modularS_{\id P}[\id] = 4\sqrt{2} \sinh(2\pi bP)\sinh(2\pi b^{-1}P) \ . 
\end{align}
The S-matrix with no operator inserted is
\begin{align}
\modularS_{P_1P_2}[\id] = 2 \sqrt{2} \cos(4\pi P_1 P_2) \ .
\end{align}
Under swapping the lower arguments,
\begin{align}
\modularS_{P_2P_1}[P_3] = \modularS_{P_1P_2}[P_3]
\frac{\rho_0(P_1)C_0(113)}{\rho_0(P_2)C_0(223)} \ . 
\end{align}
The S-matrix in Racah-Wigner normalization, following the notation in \cite{Post:2024itb}, is 
\begin{align}\label{hatSdef}
\hatS_{P_1P_2}[P_3]
&= \frac{1}{\rho_0(P_2)}\sqrt{ \frac{C_0(113)}{C_0(223)}}
\modularS_{P_1P_2}[P_3]
\end{align}
This is symmetric in the lower arguments,
\begin{align}
\hatS_{P_1P_2}[P_3] = \hatS_{P_2P_1}[P_3] \ , 
\end{align}
and the normalization is such that 
\begin{align}
\hatS_{\id P}[\id] = 1 \ ,
\qquad
\hatS_{P_1P_2}[\id] = \frac{2\sqrt{2}\cos(4\pi P_1 P_2)}{\sqrt{\rho_0(P_1)\rho_0(P_2)}} 
\end{align}
The orthogonality relation in Racah-Wigner normalization is
\begin{align}
\int_0^\infty dQ \rho_0(Q) \hatS_{P_1Q}[P_3]\hatS^*_{QP_2}[P_3]
&= \frac{1}{\rho_0(P_1) } \delta(P_1 - P_2)
\end{align}
and the $(ST)^3$ relation is
\begin{align}\label{stcubedRW}
\int_0^\infty dQ \, \hatS_{PQ}[P_3] \hatS_{QR}[P_3] e^{-2\pi i(h_P+h_Q+h_R-c/8)} 
&= \hatS_{PR}[P_3] \ . 
\end{align}
An integral representation of $\modularS_{P_1P_2}[P_3]$ was derived in \cite{Teschner:2013tqy}, and several alternative versions can be found in \cite{Eberhardt:2023mrq}. 

\subsection{Identities relating $\modularS$ and $\mathbb{F}$}
\newcommand{\bigj}{}

The Moore-Seiberg conditions imply relations between the S-matrix and the fusion kernel. From the 2-point function on the torus one finds
\begin{align}
\hatS_{P_1P_2}[P_3]&\int_0^\infty dP_4 \, \rho_0(P_4)
\begin{Bmatrix} P_0 & P_7 & P_3 \\ P_2 & P_2 & P_4 \end{Bmatrix}
\begin{Bmatrix} P_2 & P_2 & P_5 \\P_0 & P_7 & P_4 \end{Bmatrix}
e^{2\pi i(h_4-h_2)}\\
&= 
\int_0^\infty dP_6 \, \rho_0(P_6) 
\hatS_{P_6P_2}[P_5]
\begin{Bmatrix}P_6 & P_6 & P_5 \\ P_0 & P_7 & P_1 \end{Bmatrix}
\begin{Bmatrix} P_0 & P_7 & P_3 \\P_1 & P_1 & P_6 \end{Bmatrix}
e^{i\pi(h_0+h_7 - h_5)}\notag
\end{align}
A special case of this relation is \cite[(2.65)]{Eberhardt:2023mrq}
\begin{align}\label{SfromF}
\hatS^*_{P_1P_2}[P_3]
&= \int_0^\infty dP_4 \rho_0(P_4) \begin{Bmatrix}P_1&P_1&P_3\\P_2&P_2&P_4\end{Bmatrix}(\mathbb{B}_{P_4}^{P_1P_2})^2
\end{align}
with $\mathbb{B}_{P_4}^{P_1P_2} = e^{i\pi(h_4-h_1-h_2)}$. 
The irrational Verlinde formula derived by Post and Tsiares \cite{Post:2024itb} is
\begin{align}\label{ptformula}
\begin{Bmatrix}P_1 & P_1 &P_3\\P_2&P_2&P_4\end{Bmatrix}
&= \int_0^\infty dP \, 
\hatS_{P_1P}[P_3] \modularS_{P_4 P}[\id] \hatS_{PP_2}^*[P_3]
\end{align}

\addcontentsline{toc}{section}{References}
\bibliographystyle{utphys}
{\small
\input{ctv.bbl}
}

\end{document}

%% file: ctv.bbl
\providecommand{\href}[2]{#2}\begingroup\raggedright\endgroup